\documentclass[usenatbib]{mn2e}
\usepackage[dvips]{graphicx}
\usepackage{multirow}
\usepackage{amsmath}

\def\CN2{\mbox{$C_N^2$}}
\def\tauO{\mbox{$\tau_{0}$}}
\def\thetaO{\mbox{$\theta_{0}$}}
\def\see{\mbox{$\varepsilon$}}

\title[MOSE: surface layer atmospherical parameters]{MOSE: optical turbulence and atmospherical parameters operational forecast at ESO 
ground-based sites. \\
II: atmospherical parameters in the surface layer [0-30] m.}
\author[F. Lascaux et al.]{F. Lascaux$^{1}$\thanks{E-mail:
     lascaux@arcetri.astro.it; masciadri@arcetri.astro.it}, E. Masciadri$^1$\footnotemark[1] and L. Fini$^1$ \\ 
$^1$INAF Osservatorio Astrofisico di Arcetri, Largo Enrico Fermi 5, I-501 25 Florence, Italy}

\begin{document}
\newcommand{\cn}{$C_N^2$}
\label{firstpage}
\date{Accepted 2013 ??? ??, Received 2013 ??? ??; in original form
2013 ??? ??}  
\pagerange{\pageref{firstpage}--\pageref{lastpage}}
\pubyear{2013}
\maketitle
\begin{abstract}
This article is the second of a series of articles aiming at proving the feasibility of the forecast of all the most 
relevant classical atmospherical parameters for astronomical applications (wind speed and direction, temperature, 
relative humidity) and the optical turbulence ($\CN2$ and the derived astro-climatic parameters like seeing $\see$, isoplanatic angle $\thetaO$, 
wavefront coherence time $\tauO$...). 
This study is done in the framework of the MOSE project, and focused above the two ESO ground-bases sites of Cerro Paranal and 
Cerro Armazones. 
In this paper we present the results related to the Meso-Nh model ability in reconstructing the surface layer atmospherical parameters 
(wind speed intensity, wind direction and absolute temperature, [0-30]~m a.~g.~l.).
The model reconstruction of all the atmospherical parameters in the surface layer is very satisfactory. 
For the temperature, at all levels, the RMSE (Root Mean Square Error) is inferior to 1$^{\circ}$C. 
For the wind speed, it is $\sim$2~m$\cdot$s$^{-1}$, 
and for the wind direction, it is in the range [38-46$^{\circ}$], at all levels, that corresponds to a RMSE$_{relative}$ in a range [21-26\%].
If a filter is applied for the wind direction 
(the winds inferior to 3~m$\cdot$s$^{-1}$ are discarded from the computations), the wind direction RMSE is in the range [30-41$^{\circ}$], i.e. 
a RMSE$_{relative}$ in the range [17-23\%].
The model operational forecast of the surface layer atmospherical parameters is suitable for different applications, among others: thermalization of the dome
using the reconstructed temperature, hours in advance, of the beginning the night; knowing in advance the main direction which the strong winds 
will come from during the night could allow the astronomer to anticipate the occurrence of a good/bad seeing night, and plan the observations 
accordingly; preventing adaptive secondary mirrors shake generated by the wind speed.
\end{abstract}
\section{Introduction}
This paper is the second part of a general study about the feasibility of the forecast of meteorological parameters and optical turbulence 
at ESO sites (Cerro Paranal and Cerro Armazones) in the framework of the MOSE project (MOdeling ESO Sites). 
The MOSE project, and the first results obtained for the vertical stratification of different atmospheric parameters, are presented in a joint 
paper \citep{masciadri13}.
The reader can refer to this joint paper to have more details about MOSE.
We only recall here that the MOSE project aims at proving the feasibility of the forecast of the most relevant classical atmospherical parameters
for astronomical applications (wind speed intensity and direction, temperature, relative humidity) and the optical turbulence OT ($\CN2$ profiles)
with the integrated astro-climatic parameters derived from the $\CN2$ i.e. the seeing ($\varepsilon$),
the isoplanatic angle ($\thetaO$), the wavefront coherence time ($\tauO$) above the two ESO sites of Cerro Paranal
(site of the Very Large Telescope - VLT) and Cerro Armazones (site selected for the European Extremely Large Telescope - E-ELT).\\
The final outcome of the project is to investigate the opportunity to implement an automatic system for the
forecast of these parameters at the VLT Observatory at Cerro Paranal and at the E-ELT Observatory at Cerro Armazones.\\ \\
The Meso-Nh model has already successfully been used to investigate some atmospherical parameters above sites of interest for the astronomy: 
at Roque de los Muchachos (near ground temperature, \citet{masciadri01b});
at San Pedro Martir (wind speed profiles, \citet{masciadri01a});
at Cerro Paranal in Chile and Maidanak in Uzbekistan (near ground wind speed, \citet{masciadri03}); 
in Antarctica (wind speed and temperature profiles at Dome C, \citet{lascaux09});
at Mount Graham, Arizona (wind speed vertical distribution, \citet{hagelin10} and use of the Meso-Nh wind speed for wavefront coherence time reconstruction, 
\citet{hagelin11}). 
In the present study, we focus our analysis on the model ability in reconstructing 
the meteorological surface parameters (temperature, wind speed and direction), 
from the ground up to 30~m a.~g.~l, at Cerro Paranal and Cerro Armazones.
In Section 2, we describe the numerical set-up used for the mesoscale simulations.
In Section 3, we present the data-set (instruments and measurements) and the statistical parameters used to evaluate the performances of the model.
In Section 4, we evaluate the overall statistical performances of the model.
In Section 5, we look at the single nights temporal evolutions of the meteorological surface forecasted parameters.
In Section 6, we look more closely at the performances of the model for the individual nights.
In Section 7, we discuss the usefulness of the correlation coefficient in our context.
Conclusions are drawn in Section 8.
%
\section{Model configuration}                 
\begin{figure*}
\begin{center}
\begin{tabular}{c}
\includegraphics{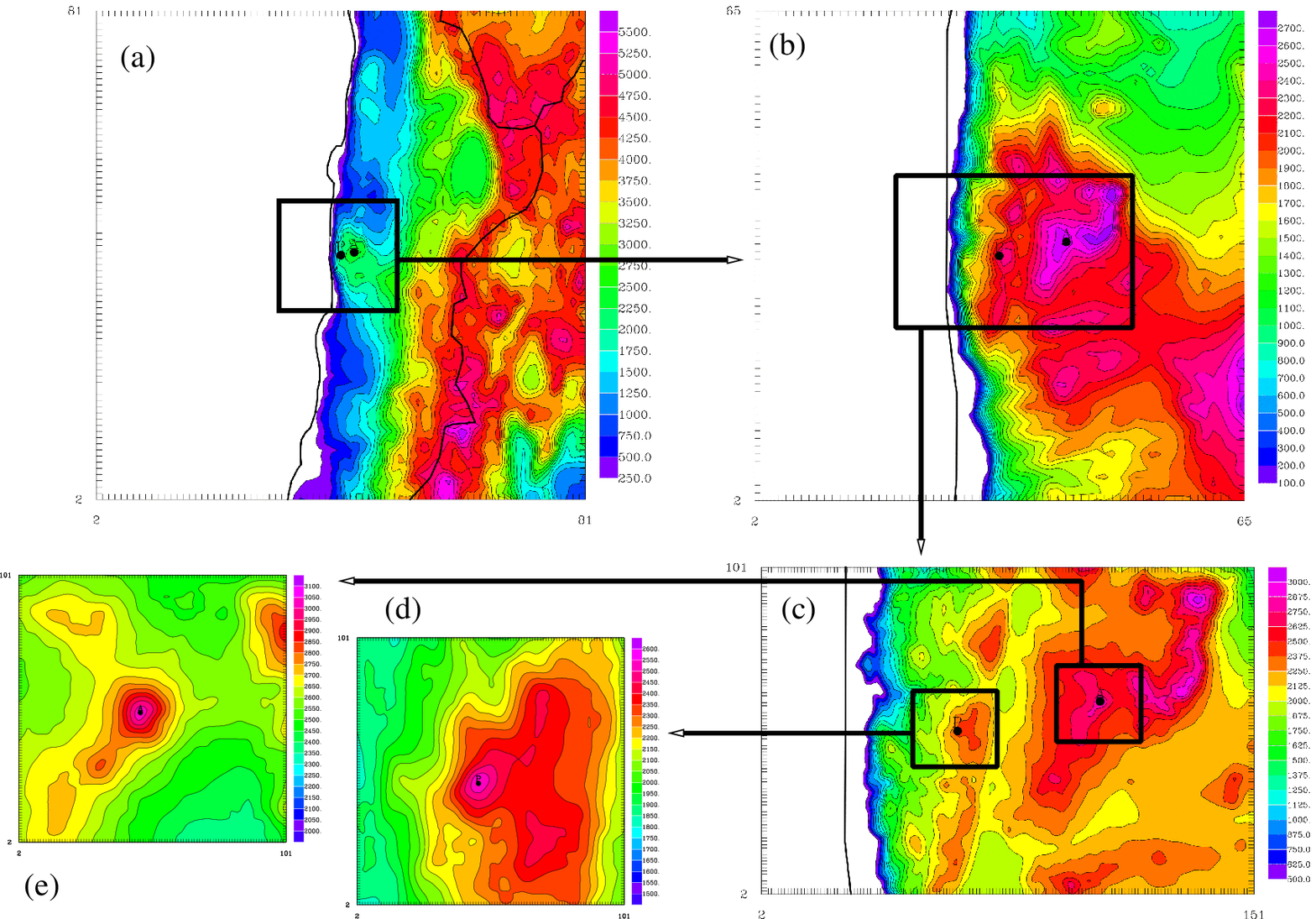}
\end{tabular}
\end{center}
\caption[oro]{\label{fig:oro} Orography (altitude in m) of the region of interest
as seen by the Meso-Nh model (polar stereographic projection)
for all the imbricated domains of the grid-nested configuration.
{\bf (a)} Domain 1 (orographic data from GTOPO),
{\bf (b)} Domain 2 (orographic data from GTOPO),
{\bf (c)} Domain 3 (orographic data from ISTAR),
{\bf (d)} Domain 4 (orographic data from ISTAR),
{\bf (e)} Domain 5 (orographic data from ISTAR),
A dot stands for Cerro Armazones and P dot stands for Cerro Paranal.
See Table \ref{tab:config} for the specifications of the domains (number of grid-points, domain extension, horizontal resolution).
The reader nay notice that domain (a), (b) and (c) are identical to the domains (a), (b) and (c) of Figure 1 of Masciadri et al. (2013).
The innermost domains (d) and (e) are only relevant in this present paper since we also present the results from simulations performed 
with these very high horizontal resolutions (${\Delta}X$~=~100~m).}
\end{figure*}

\begin{table*}
 \centering
  \caption{Geographic coordinates of the two sites investigated in this paper
           (Cerro Paranal and Cerro Armazones). The altitude is in meter. $\Delta$h represents the difference
between the Meso-Nh ground altitude and the real ground altitude.}
  \begin{tabular}{|c|r|r|c|c|c|}
   \hline
 \multirow{2}{*}{SITE}  & \multirow{2}{*}{LATITUDE}  & \multirow{2}{*}{LONGITUDE}  & \multicolumn{2}{|c|}{MESO-NH}      & MEASURED    \\
                        &                            &                             & \multicolumn{2}{|c|}{GROUND ALTITUDE (m)} & GROUND ALTITUDE (m) \\
                        &                            &                     & $\Delta$X=500 m & $\Delta$X=100 m &           \\
  \hline
 Cerro Paranal  $^*$    & 24$^{\circ}$37'33.117"S  & 70$^{\circ}$24'11.642"W & 2478 ($\Delta$h=156~m)& 2545 ($\Delta$h=89~m) & 2634 \\  
 Cerro Armazones$^{**}$ & 24$^{\circ}$35'21"S  & 70$^{\circ}$11'30"W & 2901 ($\Delta$h=164~m) & 3010 ($\Delta$h=55~m)& 3065 \\
   \hline
 \multicolumn{4}{|l|}{$^*$ {\scriptsize UT1 facility unit coordinates.}} \\
 \multicolumn{4}{|l|}{$^{**}$ {\scriptsize GPS measurement by ESO.}} \\
  \end{tabular}
 \label{tab0}
\end{table*}
All the numerical simulations of the nights presented in this study have been performed with the mesoscale numerical weather model
 Meso-Nh\footnote{$http$:$//mesonh.aero.obs$-$mip.fr/mesonh/$} \citep{lafore98}.
The model has been developed by the Centre National des Recherches M\'et\'eorologiques (CNRM) and Laboratoire d'A\'ereologie (LA)
from Universit\'e Paul Sabatier (Toulouse).
The Meso-Nh model can simulate the temporal evolution of three-dimensional meteorological
parameters over a selected finite area of the globe.
We refer the reader to Masciadri et al. (2013), Sec.~3.3, for the general model configuration and the physical packages. 
We just recall here the use of the grid-nesting technique \citep{stein00}, that consists in using different imbricated domains
of the Digital Elevation Models (DEM i.e orography) extended on smaller and smaller surfaces, with increasing horizontal
resolution but with the same vertical grid.
In this study we use two different configurations.
The first grid-nesting configuration employed three domains (Fig.~\ref{fig:oro} and
Table~\ref{tab:config}) and the innermost resolution is ${\Delta}X$~=~500~m.
The second configuration is made of five imbricated domains, the first same three as the previous configuration, and other two
centered at both Paranal and Armazones sites, with a horizontal resolution of ${\Delta}X$~=~100~m (all domains of Fig.~\ref{fig:oro}
and Table~\ref{tab:config}).
One can notice that, using these configurations, we are able to do the forecast at both sites simultaneously.
The orographic DEMs we used for this project are the GTOPO\footnote{$http$:$//www1.gsi.go.jp/geowww/globalmap$-$gsi/gtopo30/gtopo30.html$}
with an intrinsic horizontal resolution of 1~km (used for the domains 1 and 2) and the ISTAR with an intrinsic
horizontal resolution of 0.5~km (used for the domain 3).
Along the z-axis we have 62 levels distributed as follows: a first vertical grid point equal to 5~m,
a logarithmic stretching of 20~$\%$ up to 3.5~km above the ground, and an almost constant vertical grid size of $\sim$600~m up to 23.8~km.\\
All the simulations done for the analyses discussed in this paper, were initialized the day before at 18~UT
and forced every 6 hours with the analyses from the ECMWF,
and finished at 09~UT of the simulated day (for a total duration of 15 hours).
The statistics is computed only during night local time, from 00 UT to 09 UT.
\begin{table*}
\centering
\caption{Meso-Nh model configurations (Table extracted from Masciadri et al., 2013). In the second column the  horizontal resolution $\Delta$X, in the third column the number of grid points and in the fourth column the horizontal surface covered by the model domain.}
\begin{tabular}{cccc}
\hline
\multirow{2}{*}{Domain} & \multirow{2}{*}{$\Delta$X (km)} & \multirow{2}{*}{Grid Points}    & Domain size \\
                        &                                 &                                 & (km)        \\ 
\hline
Domain 1     & 10             &  80$\times$80  &  800$\times$800       \\
Domain 2     &  2.5           &  64$\times$64  &  160$\times$160       \\
Domain 3     &  0.5           & 150$\times$100 &   75$\times$50        \\
Domain 4     &  0.1           & 100$\times$100 &   10$\times$10        \\
Domain 5     &  0.1           & 100$\times$100 &   10$\times$10        \\
\hline
\end{tabular}
\label{tab:config}
\end{table*}
%
\section{Observations data-set and statistical tools}
{\bf Site(s)}: Cerro Paranal and Cerro Armazones; \\
{\bf Instruments}: Automatic Weather Station (AWS) and masts; \\
{\bf Parameters investigated}: wind speed, wind direction, absolute temperature;\\
{\bf Number of nights}: 20 nights in summer 2007 (cf. Table~\ref{tab:annexe_20n} for a complete list);\\
{\bf Atmospheric region investigated}: surface layer ([0,30~m], cf Table~\ref{tab:ari}). \\ \\
\noindent

\begin{table*}
 \centering
 \caption{List of the 20 simulated nights with available meteorological observations near the surface. These
  20 nights were used for the validation of the model meteorological performances.\label{tab:annexe_20n}}
 \begin{tabular}{|ccccc|}
 \hline
 \multicolumn{5}{|c|}{Simulated Nights - meteorological surface observations} \\
 \hline
 2007-11-02 & 2007-11-03 & 2007-11-04 & 2007-11-05 & 2007-11-06 \\
 2007-11-07 & 2007-11-11 & 2007-11-13 & 2007-11-14 & 2007-11-17 \\
 2007-11-18 & 2007-11-20 & 2007-11-27 & 2007-11-28 & 2007-11-29 \\
 2007-11-30 & 2007-12-20 & 2007-12-22 & 2007-12-23 & 2007-12-24 \\
 \hline
 \end{tabular}
\end{table*}
\begin{table*}
 \centering
 \caption{Altitudes of observation for all meteorological parameters investigated near the surface, at
          Cerro Paranal and Cerro Armazones.\label{tab:ari}}
 \begin{tabular}{|c|c|c|}
 \hline
 Parameters & Cerro Paranal & Cerro Armazones \\
 \hline
 Wind speed & 10~m / 30~m   & 2~m / 11~m / 20~m / 28~m \\
 \hline
 Wind direction & 10~m / 30~m   & 2~m / 11~m / 20~m / 28~m \\
 \hline
 Temperature  &  2~m / 30~m   & 2~m / 11~m / 20~m / 28~m \\
 \hline
 \end{tabular}
\end{table*}
During the MOSE project, our team have studied data from 20 nights of the 
PAR2007 campaign \citep{dali10}, which have been used for the model calibration of the optical turbulence 
(results will be presented in a forthcoming paper). 
We decided to use these nights also for the meteorological forecast validation. 
However, not all the nights of the PAR2007 campaign have meteorological measurements 
available at both sites contemporaneously, so we completed the data sample with other nights during the same period (November-December 2007) for which 
observations were available at both Cerro Paranal and Cerro Armazones. 
The 20 nights are listed in Table~\ref{tab:annexe_20n}.
At Cerro Paranal, observations of meteorological parameters near the surface come from an automated weather station
(AWS) and a 30~m high mast including a number of sensors at different heights. Both instruments are part
of the VLT Astronomical Site Monitor \citep{sandrock99}.
Absolute temperature data are available at 2~m and 30~m above the
ground. Wind speed data are available at 10~m and 30~m above the ground (Table~\ref{tab:ari}).
At Cerro Armazones, observations of the
meteorological parameters near the ground surface come from the Site Testing Database \citep{schoeck09}, more precisely
from an AWS and a 30~m tower (with temperature sensors and sonic anemometers). Data on temperature and
wind speed are available at 2~m, 11~m, 20~m and 28~m above the ground (Table~\ref{tab:ari}). At 2~m (Armazones) temperature
measurements from the AWS and the sonic anemometers are both available but we considered only those from
the tower (accuracy of 0.10 $^{\circ}$C) \citep{skidmore07}.
Those from the AWS are not reliable because of some drift effects (T. Travouillon, private communication).
Wind speed observations are taken from the AWS (at 2 m) and from the sonic
anemometers of the tower (at 11~m, 20~m and 28 m).
The outputs are sampled with a temporal frequency of 1 minute.\\ \\
\noindent
To estimate the statistical model reliability in reconstructing the main meteorological parameters we used the averaged values plus
two statistical operators: the bias and the root mean square error (RMSE).
\begin{equation}
BIAS = \sum_{i=1}^{N}\frac{({\Delta}_i)}{N}
\label{eq:bias}
\end{equation}
\begin{equation}
RMSE = \sqrt{\sum_{i=1}^{N}\frac{({\Delta}_i)^2}{N}}
\label{eq:rmse}
\end{equation}
with ${\Delta}_i=Y_i-X_i$ 
where $X_i$ are the individual observations, $Y_i$ the individual simulations parameters calculated at the same time and $N$ is
the number of times for which a couple ($X_i$,$Y_i$) is available with both $X_i$ and $Y_i$ different from zero.
Because the wind direction is a circular variable, we define ${\Delta}_i$ for the wind direction as:
\begin{equation}
\Delta_i = \begin{cases}
Y_i-X_i     & \text{if $\left | Y_i-X_i  \right | \leq 180{^{\circ}}$} \\
Y_i-X_i-360{^{\circ}} & \text{if $Y_i-X_i >  180{^{\circ}}$} \\
Y_i-X_i+360{^{\circ}} & \text{if $Y_i-X_i < -180{^{\circ}}$}
\end{cases}
\label{eq:deltai}
\end{equation}
We decided not to analyse the correlation coefficient, and
a dedicated discussion on its uselessness for our study will be done in Section~\ref{sec:cc}.
At the same time, due to the fact that we are interested in investigating the model ability in forecasting a parameter
and not only in characterizing it, it is important to investigate also the correlation observations/simulations
calculated night by night and not only in statistical terms.
We also report in every scattered plot the slope of a regression line passing by the origin.
%
\section{Overall statistical model performances}
We first proceed in this section to an overall statistical study of the model performances in reconstructing temperature,
wind speed and wind direction near the ground at
both sites (Cerro Paranal and Cerro Armazones).
All the statistical computations are made on the whole sample of data covering 20 nights of observations (Table~\ref{tab:annexe_20n}).
\begin{figure*}
\begin{center}
\begin{tabular}{c}
\includegraphics[width=0.65\textwidth]{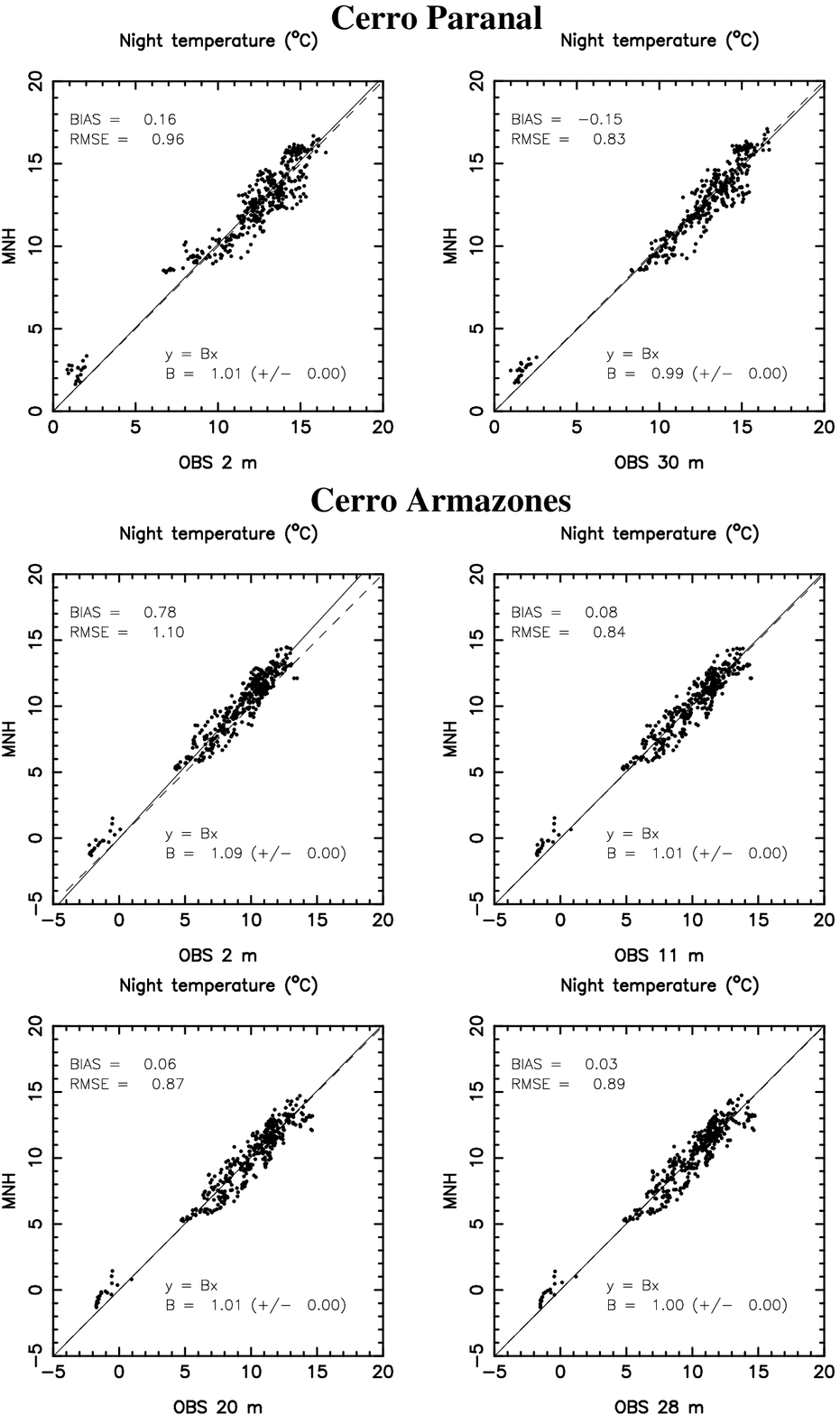}
\end{tabular}
\end{center}
\caption[Surface_temp_cloud]{\label{fig:Surface_temp_cloud} Scattered plot of Meso-Nh temperature against observations,
at 2~m and 30~m at Cerro Paranal (top figures); at 2~m, 11~m, 20~m and 28~m at Cerro Armazones (central and bottom figures).
Every point represents the average over an interval of 30 minutes. The dashed lines represent the ideal $y=x$ relationship. 
The thin straight lines are the regression lines of the sample $y=B{\cdot}x$, passing by the origin. See Table~\ref{tab:br_temp}. }
\end{figure*}
\begin{figure*}
\begin{center}
\begin{tabular}{c}
\includegraphics[width=0.65\textwidth]{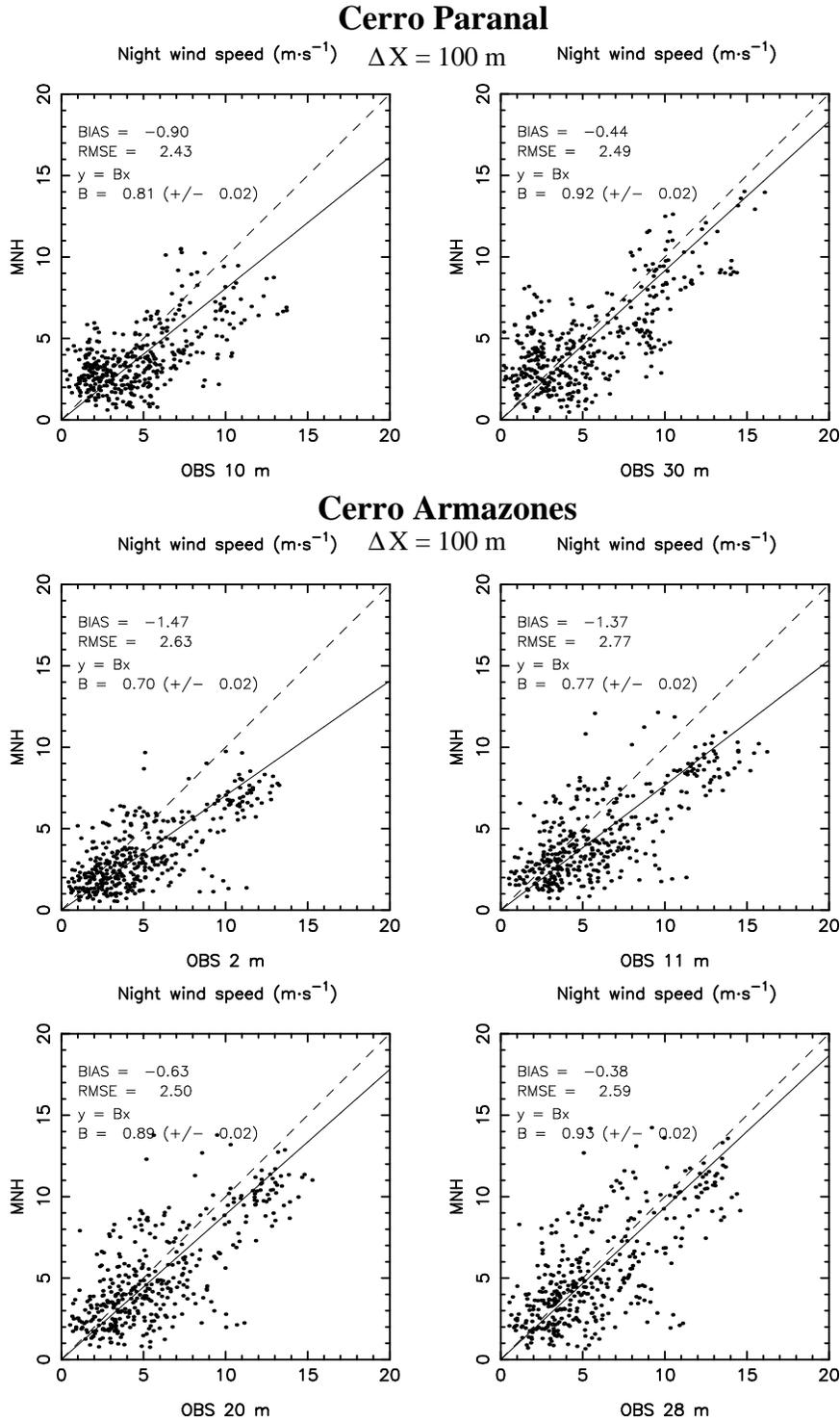}
\end{tabular}
\end{center}
\caption[Surface_ws_cloud_5mod]{\label{fig:Surface_ws_cloud_5mod} Scattered plot of Meso-Nh wind speed
against observations, at 10~m and 30~m at Cerro Paranal (top figures);
at 2~m, 11~m, 20~m and 28~m at Cerro Armazones (central and bottom figures). Meso-Nh configuration is the $\Delta$X~=~100~m configuration.
Every point represents the average over an interval of 30 minutes. The dashed lines represent the ideal $y=x$ relationship. 
The thin straight lines are the regression lines of the sample $y=B{\cdot}x$, passing by the origin. See Table~\ref{tab:br_ws_5mod}.}
\end{figure*}
\begin{figure*}
\begin{center}
\begin{tabular}{c}
\includegraphics[width=0.65\textwidth]{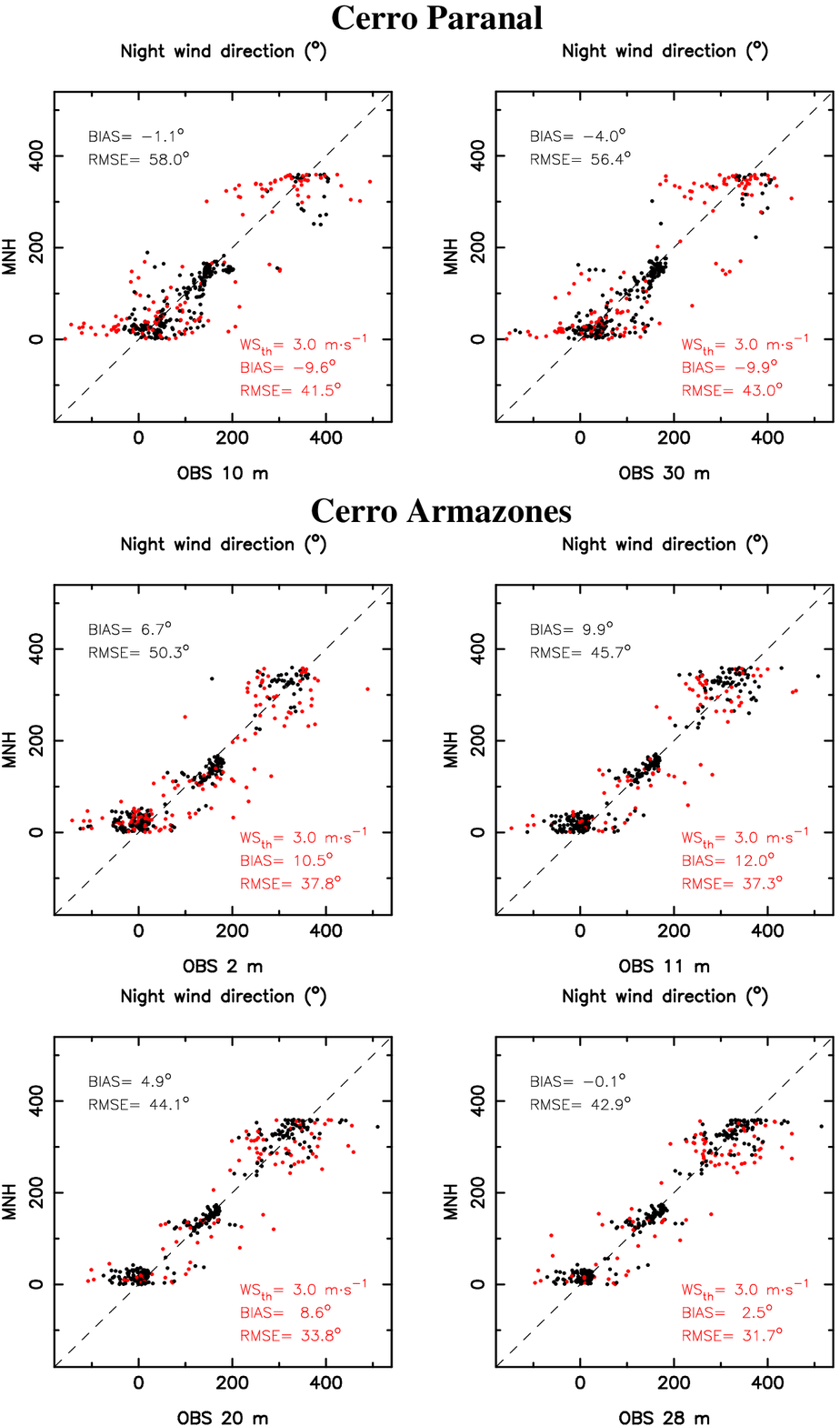}
\end{tabular}
\end{center}
\caption[Surface_wd_cloud]{\label{fig:Surface_wd_cloud} Scattered plot of Meso-Nh wind direction against observations,
at 10~m and 30~m at Cerro Paranal (top figures);
at 2~m, 11~m, 20~m and 28~m at Cerro Armazones (central and bottom figures).
Every point represents the average over an interval of 30 minutes.
The red dots correspond to points for which the observed wind was inferior to $WS_{th}$~=~3~m$\cdot$s$^{-1}$.
The corresponding values of bias and RMSE 
for the sample without these dots are reported in red in the bottom right of each plot.
Notice that no regression line is calculated in the case of wind direction comparisons. See Table~\ref{tab:br_wd}.}
\end{figure*}
Fig.~\ref{fig:Surface_temp_cloud} displays the scattered plots of the model computed absolute temperature 
against the observed absolute temperature, for both sites,
and at every altitudes for which observations were available (see Table~\ref{tab:ari}).
Fig.~\ref{fig:Surface_ws_cloud_5mod} displays the scattered plots of the model computed wind speed
(with the $\Delta$X~=~100 m configuration) against the observed wind speed, for both sites,
and at every altitudes for which observations were available (see Table~\ref{tab:ari}).
Finally, Fig.~\ref{fig:Surface_wd_cloud} displays the scattered plots of the model computed wind direction 
against the observed wind direction, for both sites,
and at every altitudes for which observations were available (see Table~\ref{tab:ari}).
All the bias and RMSE values reported in these figures are also summarized in Tables \ref{tab:br_temp}, \ref{tab:br_ws}, \ref{tab:br_ws_5mod}, 
\ref{tab:br_wd} and \ref{tab:br_wd_wsth}.
In all the aforementioned figures, every point is the result of a 30-minutes average for both observations and model.
\begin{table*}
 \centering
 \caption{Near surface temperature bias and RMSE (Meso-Nh minus Observations), using the data with re-sampling
 average over 30 min. See Fig.~\ref{fig:Surface_temp_cloud}. \label{tab:br_temp}}
 \begin{tabular}{|c|c|c||c|c|c|c|}
 \multicolumn{7}{c}{Absolute temperature ($^{\circ}$C)} \\
 \cline{2-7}
 \multicolumn{1}{c}{ } & \multicolumn{2}{|c||}{PARANAL} & \multicolumn{4}{|c|}{ARMAZONES} \\
 \multicolumn{1}{c}{ } & \multicolumn{1}{|c}{2 m} & \multicolumn{1}{c||}{30 m} &
 \multicolumn{1}{|c}{2 m} & \multicolumn{1}{c}{11 m} & \multicolumn{1}{c}{20 m} &
 \multicolumn{1}{c|}{28 m} \\
\hline
 BIAS     & 0.16  & -0.15  &  0.78  & 0.08   &  0.06 &  0.03 \\
 RMSE     & 0.96  & 0.83   &  1.10  & 0.84   & 0.87  & 0.89  \\
 \hline
 \end{tabular}
\end{table*}
\begin{table*}
 \centering
 \caption{Near surface wind speed, bias and RMSE (Meso-Nh with the
  standard configuration - maximum $\Delta$X~=~500 m -  minus Observations). \label{tab:br_ws}}
 \begin{tabular}{|c|c|c||c|c|c|c|}
 \multicolumn{7}{c}{Wind speed (m$\cdot$s$^{-1}$) - $\Delta$X~=~500 m} \\
 \cline{2-7}
 \multicolumn{1}{c}{ } & \multicolumn{2}{|c||}{PARANAL} & \multicolumn{4}{|c|}{ARMAZONES} \\
 \multicolumn{1}{c}{ } & \multicolumn{1}{|c}{10 m} & \multicolumn{1}{c||}{30 m} &
 \multicolumn{1}{|c}{2 m} & \multicolumn{1}{c}{11 m} & \multicolumn{1}{c}{20 m} &
 \multicolumn{1}{c|}{28 m} \\
\hline
 BIAS     & -2.17  & -1.07  &  -3.58  & -3.31 & -2.33  &  -2.00 \\
 RMSE     & 3.14  & 2.62   &  4.54  & 4.74   & 4.20  & 4.15  \\
 \hline
 \end{tabular}
\end{table*}
\begin{table*}
 \centering
 \caption{Near surface wind speed, bias and RMSE (Meso-Nh with the
  high horizontal resolution configuration - maximum $\Delta$X~=~100 m  - minus Observations). See Fig.~\ref{fig:Surface_ws_cloud_5mod}. \label{tab:br_ws_5mod}}
 \begin{tabular}{|c|c|c||c|c|c|c|}
 \multicolumn{7}{c}{Wind speed (m$\cdot$s$^{-1}$) - $\Delta$X~=~100 m} \\
 \cline{2-7}
 \multicolumn{1}{c}{ } & \multicolumn{2}{|c||}{PARANAL} & \multicolumn{4}{|c|}{ARMAZONES} \\
 \multicolumn{1}{c}{ } & \multicolumn{1}{|c}{10 m} & \multicolumn{1}{c||}{30 m} &
 \multicolumn{1}{|c}{2 m} & \multicolumn{1}{c}{11 m} & \multicolumn{1}{c}{20 m} &
 \multicolumn{1}{c|}{28 m} \\
\hline
 BIAS     & -0.90  & -0.44  &  -1.47  & -1.37 & -0.63  &  -0.38 \\
 RMSE     & 2.43  & 2.49   &  2.63  & 2.77   & 2.50  & 2.59  \\
 \hline
 \end{tabular}
\end{table*}
\begin{table*}
 \centering
 \caption{Near surface wind direction, bias and RMSE (Meso-Nh with the
  standard configuration - maximum $\Delta$X~=~500 m -  minus Observations).  See Fig.~\ref{fig:Surface_wd_cloud}.\label{tab:br_wd}}
 \begin{tabular}{|c|c|c||c|c|c|c|}
 \multicolumn{7}{c}{Wind direction ($^{\circ}$)} \\
 \cline{2-7}
 \multicolumn{1}{c}{ } & \multicolumn{2}{|c||}{PARANAL} & \multicolumn{4}{|c|}{ARMAZONES} \\
 \multicolumn{1}{c}{ } & \multicolumn{1}{|c}{10 m} & \multicolumn{1}{c||}{30 m} &
 \multicolumn{1}{|c}{2 m} & \multicolumn{1}{c}{11 m} & \multicolumn{1}{c}{20 m} &
 \multicolumn{1}{c|}{28 m} \\
\hline
 BIAS     & -1.1  & -4.0  &  6.7  & 9.9 & 4.9  &  -0.1 \\
 RMSE     & 58.0  & 56.4   &  50.3  & 45.7   & 44.1  & 42.9  \\
 \hline
 \end{tabular}
\end{table*}
\begin{table*}
 \centering
 \caption{Near surface wind direction, bias and RMSE (Meso-Nh with the
  standard configuration - maximum $\Delta$X~=~500 m -  minus Observations).  See Fig.~\ref{fig:Surface_wd_cloud}.
  Here the points for which the wind speed was inferior to 3~m$\cdot$s$^{-1}$ were discarded in the computations of the
  bias and the RMSE (red dots of Fig.~\ref{fig:Surface_wd_cloud}). \label{tab:br_wd_wsth}}
 \begin{tabular}{|c|c|c||c|c|c|c|}
 \multicolumn{7}{c}{Wind direction ($^{\circ}$)} \\
 \cline{2-7}
 \multicolumn{1}{c}{ } & \multicolumn{2}{|c||}{PARANAL} & \multicolumn{4}{|c|}{ARMAZONES} \\
 \multicolumn{1}{c}{ } & \multicolumn{1}{|c}{10 m} & \multicolumn{1}{c||}{30 m} &
 \multicolumn{1}{|c}{2 m} & \multicolumn{1}{c}{11 m} & \multicolumn{1}{c}{20 m} &
 \multicolumn{1}{c|}{28 m} \\
\hline
 BIAS     & -9.6  & -9.9  & 10.5  &12.0 & 8.6  &   2.5 \\
 RMSE     & 41.5  & 43.0   &  37.8  & 37.3   & 33.8  & 31.7  \\
 \hline
 \end{tabular}
\end{table*}
\subsection{Absolute temperature and wind speed}
The results in terms of absolute temperature are particularly impressive. Considering both sites, and every levels, the bias is very
small: it goes from 0.03$^{\circ}$C
at 28~m at Cerro Armazones (smallest value) to 0.78$^{\circ}$C at 2~m at Cerro Armazones (largest value), well below 1$^{\circ}$C.
Even the RMSE is almost always inferior to 1$^{\circ}$C, which is very satisfying (the largest value of 1.10$^{\circ}$C is encountered near the ground at 2~m 
at Cerro Armazones). 
These are very interesting values considering that the typical values for the temperature (and for the period considered with this sample, i.e. summer) 
are between 11$^{\circ}$C and 13$^{\circ}$C at Cerro Paranal and between 8$^{\circ}$C and 10$^{\circ}$C at Cerro Armazones. \\ \\
\noindent
As for the wind speed near the surface, we can see (Table~\ref{tab:br_ws} - the corresponding scattered plot is not shown here) that the model 
with the standard $\Delta$X~=~500~m configuration, underestimates the observed wind velocity, especially
at the first level of observations (2~m).
The bias goes from 1.07~m$\cdot$s$^{-1}$ at 30~m at Cerro Paranal to 3.58~m$\cdot$s$^{-1}$ at 2~m at Cerro Armazones.
The RMSE can reach values as large as 4.74~m$\cdot$s$^{-1}$ (at 11~m at Cerro Armazones).
These values are not important in absolute terms but they are not negligible in relative terms.
Indeed, the wind speed has an average value during the night of around 5~m$\cdot$s$^{-1}$ at 10~m at Cerro Paranal and at 2~m at Cerro Armazones.
At the higher elevations at Cerro Paranal it is around 5-6~m$\cdot$s$^{-1}$, and at Cerro Armazones it is around 6-7~m$\cdot$s$^{-1}$.
This mean that the relative error can be as large as 50\% for some cases.
This understimation can be explained by the relative smoothing
of the orography with $\Delta$X~=~500~m that could generate a weaker than observed wind speed over the mountainous peaks (like
at Cerro Paranal and Cerro Armazones).
One way to overcome this situation is to increase the horizontal resolution. 
Fig.~\ref{fig:Surface_ws_cloud_5mod}  shows results obtained with the $\Delta$X~=~100~m configuration. 
Table~\ref{tab:br_ws_5mod} summarized these results.
There is an evident improvement of the wind speed reconstructed over Cerro Paranal and Cerro Armazones by the model of a factor $\sim$2.
The bias strongly reduces at the higher levels (0.38~m$\cdot$s$^{-1}$ at 28~m at Cerro Armazones and 0.44~m$\cdot$s$^{-1}$ at 30~m at Cerro Paranal).
Its largest value is only 1.47~m$\cdot$s$^{-1}$ at 2~m at Cerro Armazones).
The RMSE is halved, from a maximum value of 4.74~m$\cdot$s$^{-1}$ (at 11~m at Cerro Armazones) with the $\Delta$X~=~500~m configuration to
a maximum value of 2.77~m$\cdot$s$^{-1}$  (at 11~m at Cerro Armazones) with the $\Delta$X~=~100~m configuration, a much more satisfactory result.
The relative error that reached in some cases $\sim$50\% with the $\Delta$X~=~500~m configuration, is now only about 20\% with the $\Delta$X~=~100~m configuration.
It is worth noticing that in the $\Delta$X~=~100~m configuration (not shown here), the simulated absolute temperature remains very good.
This tells us, in case one uses the $\Delta$X~=~100~m configuration, satisfactory results for both temperature and wind speed are obtained. 
\subsection{Wind direction}
\label{sec:wd_ws}
To our knowledge, few studies were aimed at deeply analysing the quality of the forecast of the wind direction by mesoscale model.
By construction, the values of the wind direction bias are in the range [-180$^{\circ}$,180$^{\circ}$], that correspond to the maximum bias possible 
(the wind direction is a circular variable).
We can define the relative RMSE for the wind direction (as in \citet{jimenez13}) as:
\begin{equation}
RMSE_{relative}=RMSE/180^{\circ}
\label{eq:rmse_rel}
\end{equation}
because the maximum RMSE is 180$^{\circ}$.\\
For the ground-based astronomy, it is more important to know the wind direction  when the wind is strong.
When the wind velocity is weak, knowing its direction becomes irrelevant.
This is particularly true when one studies the vibrations on adaptive secondaries (see Masciadri et al., 2013, for a more
extended discussion).
Moreover several studies pointed out a dependency of the wind direction variability on the inverse of the wind speed \citep{joffre88,davies99,mahrt11}.
With these considerations in mind, we have analyse both the complete sample (no filter applied) and a sub-sample with the data filtered 
using a threshold of 3~m$\cdot$s$^{-1}$.
In Fig.~\ref{fig:Surface_wd_cloud} we reported the scattered plots of the wind direction.
The red dots correspond to points for which the observed wind speed was inferior to 3~m$\cdot$s$^{-1}$. 
In the top left of each plot, in black, are reported the values of the bias and RMSE for the whole sample at the corresponding level.
In the bottom right of each plot, in red, are reported the values of the bias and RMSE of the filtered sample (the red dots corresponding to an 
observed wind speed inferior to 3~m$\cdot$s$^{-1}$, are discarded from the calculations: only the black dots have been considered).
If we look at all the data without filtering the weakest winds (i.e. all the points, red and black, of Fig.~\ref{fig:Surface_wd_cloud}), the resulting bias, 
at every level, is close to zero, in the interval [0.1~-~9.9$^{\circ}$], which is impressive.
More over, the RMSE is between 43$^{\circ}$ and 58$^{\circ}$, which also means that
in other words we have a RMSE always smaller than a quadrant. 
This corresponds to a RMSE$_{relative}$ between 24\% and 32\%.\\
The scattered plots with the filtered data are in Fig.~\ref{fig:Surface_wd_cloud} (bias and RMSE values reported in red), 
and the bias and RMSE are summarized in Table~\ref{tab:br_wd_wsth}. 
The bias remain very similar (with respect to the bias with no filtering of the weak winds) and is always inferior to 12$^{\circ}$.
The RMSE is much smaller and is now between 31.7$^{\circ}$ ($RMSE_{relative}$=18\%), at 28~m at Cerro Armazones and 43$^{\circ}$ ($RMSE_{relative}$=24\%), 
at 10~m at Cerro Paranal, which represents a gain up to
around 15$^{\circ}$ (8\% for $RMSE_{relative}$) with respect to the RMSE with no filtering. 
We verified that the model is particularly sensitive to the threshold of the wind speed.
The improvements of the RMSE is therefore better with a threshold of 3~m$\cdot$s$^{-1}$ than with 2~m$\cdot$s$^{-1}$
which would provide a gain of 10$^{\circ}$ (5\% for RMSE$_{relative}$).\\
Considering the complicated orography of both sites (local peaks surrounded by mountainous terrains), and the known difficulties in forecasting the wind 
direction in such conditions, these results are very satisfactory.
They are, for example, in line with recent results obtained using another mesocale model, WRF, by \citet{jimenez13}, and even better.
Indeed, using a nesting configuration with the innermost horizontal resolution of 2~km 
(we remind the reader that we use a 500~m horizontal mesh-size in our innermost domains), 
they found a $RMSE_{relative}$ over the mountains around 28\%. One should bear in mind that the results 
of \cite{jimenez13} were obtained 
at a different site (over the northeastern Iberian peninsula), a region with complex terrain but with somehow lower mountains than Cerro Paranal 
or Cerro Armazones. Moreover, they analysed both day and night winds, whereas we considered in the present study only night winds.
The comparison is therefore done in qualitative terms only.
\section{Temporal evolutions}
Fig.~\ref{fig:Surface_temp_paranal_evol} and Fig.~\ref{fig:Surface_temp_armazones_evol} (Appendix A) 
show the temporal evolution of the average, the bias and the RMSE
(at different levels and above Cerro Paranal and Cerro Armazones, respectively) of the absolute temperature.
Looking at Fig.~\ref{fig:Surface_temp_paranal_evol} and  Fig.~\ref{fig:Surface_temp_armazones_evol} we conclude that,
above both Cerro Paranal and Cerro Armazones, we obtain excellent bias and RMSE values: the bias is well below 1$^{\circ}$C
(at some heights well inferior to 0.5$^{\circ}$C) and, even more impressive, the RMSE is basically always inferior to 1$^{\circ}$C.\\
\noindent
Fig.~\ref{fig:Surface_ws_paranal_evol} and Fig.~\ref{fig:Surface_ws_armazones_evol} show the temporal evolution of the average, the bias and the RMSE
(at different levels and above Cerro Paranal and Cerro Armazones, respectively) of the wind speed using the $\Delta$X~=~100~m configuration 
with 5 imbricated domains for Meso-Nh.
We note a general residual tendency of the model in slightly underestimating the wind speed all along the night, particularly at the first level (2~m).
But the understimation is limited and the overall results are very satisfactory.
We present the results for the $\Delta$X~=~100~m configuration only because we found that such a new configuration definitely and substantially improves 
the model performances with respect to the $\Delta$X~=~500~m configuration.
The improvement in reconstructing the wind speed near the surface is of around 50$\%$ (with respect to the $\Delta$X~=~500~m configuration, not shown here).\\
\noindent
Fig.~\ref{fig:Surface_wd_paranal_evol} and Fig.~\ref{fig:Surface_wd_armazones_evol} show the temporal evolution of the average, the bias and the RMSE 
(at different levels and above Cerro Paranal and Cerro Armazones, respectively) of the wind direction.
During the night, the absolute bias is inferior or close to 20$^{\circ}$ at Cerro Armazones, and inferior to 10$^{\circ}$ at Cerro Paranal.
For both sites, the RMSE remains inferior to 60$^{\circ}$. 
Those are excellent results that demonstrate the ability of the model in predicting 
the general wind direction. \\ \\
\noindent
More over, for all the atmospherical parameters near the surface, at all levels and at both sites, 
the general trend during the night is fairly reproduced by the model.
We want to highlight here, once again, that the first hours of the simulation (thus during the day) are not the subject of our study, which is focused on the 
reconstruction of the atmospherical parameters by the model during the night.
Studying the evolution of these parameters during the day would imply changing the model initialization, in order to begin the simulation hours before the 
current chosen time of 18~UT. \\  
%
\section{Individual nights model performances}
In this section we present the analysis done on the model performances in reconstructing the atmospheric parameters, night by night.
We computed for every single night (from the 20 nights sample of Table~\ref{tab:annexe_20n}) the bias,
RMSE between model and observations, for both sites and at every level near the surface where
observations were available, for the absolute temperature and the wind speed and direction.
We calculate the cumulative distributions for each parameter.
We have then computed the median, first and third quartiles, summarized in Tables \ref{tab:br_temp_cumdist},
\ref{tab:br_ws_cumdist}, \ref{tab:br_ws_cumdist_5dom} and \ref{tab:br_wd_cumdist}, for the temperature, the wind speed and the wind direction, respectively.
In Fig.~\ref{fig:Surface_temp_paranal_cumdist} and Fig.~\ref{fig:Surface_temp_armazones_cumdist}
are reported the cumulative distributions of bias and RMSE of the 20 nights
for the absolute temperature at Cerro Paranal (at 2~m and 30~m) and Cerro Armazones (at 2~m, 11~m, 20~m and 28~m), respectively.\\
In Fig.~\ref{fig:Surface_ws_paranal_cumdist} and Fig.~\ref{fig:Surface_ws_armazones_cumdist}
are reported the cumulative distributions of bias and RMSE of the 20 nights
for the wind speed at Cerro Paranal (at 10~m and 30~m) and Cerro Armazones (at 2~m, 11~m, 20~m and 28~m), respectively. \\
In Fig.~\ref{fig:Surface_wd_paranal_cumdist} and Fig.\ref{fig:Surface_wd_armazones_cumdist}
are reported the cumulative distributions of bias and RMSE of the 20 nights
for the wind direction at Cerro Paranal (at 10~m and 30~m) and Cerro Armazones (at 2~m, 11~m, 20~m and 28~m), respectively.
%
%
\begin{table*}
 \centering
 \caption{Near surface median bias and RMSE (Meso-Nh minus Observations), of the temperature
 from the single nights (Table~\ref{tab:annexe_20n}) values (see Fig.~\ref{fig:Surface_temp_paranal_cumdist}). In small fonts, the 1st and 3rd quartiles.
 \label{tab:br_temp_cumdist}}
 \begin{tabular}{|c|c|c||c|c|c|c|}
 \multicolumn{7}{c}{Absolute temperature ($^{\circ}$C)} \\
 \cline{2-7}
 \multicolumn{1}{c}{ } & \multicolumn{2}{|c||}{PARANAL} & \multicolumn{4}{|c|}{ARMAZONES} \\
 \multicolumn{1}{c}{ } & \multicolumn{1}{|c}{2 m} & \multicolumn{1}{c||}{30 m} &
 \multicolumn{1}{|c}{2 m} & \multicolumn{1}{c}{11 m} & \multicolumn{1}{c}{20 m} &
 \multicolumn{1}{c|}{28 m} \\
\hline
     & & & & & & \\
 BIAS     & $0.28_{-0.33}^{+0.57}$  & $-0.03_{-0.48}^{+0.17}$  &  $0.64_{+0.50}^{+1.12}$  & $0.04_{-0.25}^{+0.50}$  & $0.11_{-0.45}^{+0.62}$ &  $0.09_{-0.57}^{+0.70}$ \\
     & & & & & & \\
 RMSE     & $0.92_{+0.66}^{+1.08}$  & $0.64_{+0.56}^{+0.90}$   &  $0.87_{+0.76}^{+1.23}$  & $0.73_{+0.56}^{+1.02}$  & $0.85_{+0.61}^{+1.01}$ & $0.93_{+0.63}^{+1.08}$  \\
     & & & & & & \\
 \hline
 \end{tabular}
\end{table*}
\begin{table*}
 \centering
 \caption{Near surface median bias and
 RMSE (Meso-Nh minus Observations) of the wind speed, using the Meso-Nh $\Delta$X~=~500 m configuration
 from the single nights (Table~\ref{tab:annexe_20n}) values (see Fig.~\ref{fig:Surface_ws_paranal_cumdist}).
 In small fonts, the 1st and 3rd quartiles. \label{tab:br_ws_cumdist}}
 \begin{tabular}{|c|c|c||c|c|c|c|}
 \multicolumn{7}{c}{Wind speed (m$\cdot$s$^{-1}$) - $\Delta$X~=~500 m} \\
 \cline{2-7}
 \multicolumn{1}{c}{ } & \multicolumn{2}{|c||}{PARANAL} & \multicolumn{4}{|c|}{ARMAZONES} \\
 \multicolumn{1}{c}{ } & \multicolumn{1}{|c}{10 m} & \multicolumn{1}{c||}{30 m} &
 \multicolumn{1}{|c}{2 m} & \multicolumn{1}{c}{11 m} & \multicolumn{1}{c}{20 m} &
 \multicolumn{1}{c|}{28 m} \\
\hline
     & & & & & & \\
 BIAS     & $-1.86_{-2.82}^{-0.77}$  & $-0.82_{-1.92}^{+0.24}$  &  $-2.40_{-5.85}^{-1.80}$  & $-1.92_{-5.72}^{-1.15}$   & $-1.38_{-4.14}^{+0.05}$  &  $-1.27_{-2.21}^{+0.26}$ \\
     & & & & & & \\
 RMSE     & $2.25_{+1.71}^{+3.22}$   & $1.91_{+1.53}^{+2.78}$   &  $2.71_{+2.14}^{+6.38}$  & $2.31_{+1.83}^{+6.41}$   & $2.38_{+1.87}^{+5.50}$  & $2.61_{+1.91}^{+2.99}$  \\
     & & & & & & \\
 \hline
 \end{tabular}
\end{table*}
\begin{table*}
 \centering
 \caption{Near surface median bias and RMSE (Meso-Nh minus Observations)
of the wind speed, using the Meso-Nh $\Delta$X~=~100 m
configuration from the single nights (Table~\ref{tab:annexe_20n}) values
(see Fig.~\ref{fig:Surface_ws_paranal_cumdist}). In small fonts, the 1st and 3rd quartiles. \label{tab:br_ws_cumdist_5dom}}
 \begin{tabular}{|c|c|c||c|c|c|c|}
 \multicolumn{7}{c}{Wind speed (m$\cdot$s$^{-1}$) - $\Delta$X~=~100 m} \\
 \cline{2-7}
 \multicolumn{1}{c}{ } & \multicolumn{2}{|c||}{PARANAL} & \multicolumn{4}{|c|}{ARMAZONES} \\
 \multicolumn{1}{c}{ } & \multicolumn{1}{|c}{10 m} & \multicolumn{1}{c||}{30 m} &
 \multicolumn{1}{|c}{2 m} & \multicolumn{1}{c}{11 m} & \multicolumn{1}{c}{20 m} &
 \multicolumn{1}{c|}{28 m} \\
\hline
     & & & & & & \\
 BIAS  & $-0.70_{-1.59}^{+0.09}$  & $-0.55_{-1.69}^{+0.43}$  &  $-0.93_{-2.98}^{-0.27}$  & $-0.93_{-2.75}^{-0.00}$   & $-0.47_{-1.90}^{+0.29}$  &  $-0.12_{-1.78}^{+1.06}$ \\
     & & & & & & \\
 RMSE  & $1.85_{+1.39}^{+2.76}$   & $2.13_{+1.34}^{+2.95}$   &  $1.86_{+1.19}^{+3.34}$  & $2.13_{+1.48}^{+3.27}$   & $2.18_{+1.47}^{+2.52}$  & $2.05_{+1.48}^{+2.68}$  \\
     & & & & & & \\
 \hline
 \end{tabular}
\end{table*}
\begin{table*}
 \centering
 \caption{Near surface median circular bias and RMSE (Meso-Nh minus Observations), of the wind
 direction from the single nights (Table~\ref{tab:annexe_20n})
 values (see Fig.~\ref{fig:Surface_wd_paranal_cumdist}). In small fonts, the 1st and 3rd quartiles.
 \label{tab:br_wd_cumdist}}
 \begin{tabular}{|c|c|c||c|c|c|c|}
 \multicolumn{7}{c}{Wind direction ($^{\circ}$)} \\
 \cline{2-7}
 \multicolumn{1}{c}{ } & \multicolumn{2}{|c||}{PARANAL} & \multicolumn{4}{|c|}{ARMAZONES} \\
 \multicolumn{1}{c}{ } & \multicolumn{1}{|c}{10 m} & \multicolumn{1}{c||}{30 m} &
 \multicolumn{1}{|c}{2 m} & \multicolumn{1}{c}{11 m} & \multicolumn{1}{c}{20 m} &
 \multicolumn{1}{c|}{28 m} \\
\hline
     & & & & & & \\
 BIAS     & $-1.01_{-18.85}^{+26.09}$  & $-5.45_{-15.20}^{+13.54}$  &  $7.28_{-11.94}^{+16.55}$  & $6.25_{-5.42}^{+19.25}$  & $2.28_{-5.63}^{+13.78}$ &  $-4.53_{-10.85}^{+9.69}$ \\
     & & & & & & \\
 RMSE     & $45.40_{+27.93}^{+77.45}$  & $46.16_{+20.94}^{+76.97}$   &  $44.21_{+22.03}^{+61.97}$  & $40.25_{+16.54}^{+58.71}$  & $37.57_{+13.45}^{+58.13}$ & $39.44_{+14.20}^{+58.74}$  \\
     & & & & & & \\
 \hline
 \end{tabular}
\end{table*}
\begin{table*}
 \centering
 \caption{Same as Table~\ref{tab:br_wd_cumdist}, with the winds inferior to 3~m$\cdot$s$^{-1}$ discarded from the computations.
 \label{tab:br_wd_cumdist_filter}}
 \begin{tabular}{|c|c|c||c|c|c|c|}
 \multicolumn{7}{c}{Wind direction ($^{\circ}$)} \\
 \cline{2-7}
 \multicolumn{1}{c}{ } & \multicolumn{2}{|c||}{PARANAL} & \multicolumn{4}{|c|}{ARMAZONES} \\
 \multicolumn{1}{c}{ } & \multicolumn{1}{|c}{10 m} & \multicolumn{1}{c||}{30 m} &
 \multicolumn{1}{|c}{2 m} & \multicolumn{1}{c}{11 m} & \multicolumn{1}{c}{20 m} &
 \multicolumn{1}{c|}{28 m} \\
\hline
     & & & & & & \\
 BIAS     & $-1.01_{-20.22}^{+6.92} $  & $-8.55_{-24.43}^{+0.27} $  &  $7.95_{-19.30}^{+25.90}$  & $8.10_{-5.15}^{+15.78}$  & $2.38_{-7.86}^{+12.25}$ &  $-2.48_{-12.34}^{+10.45}$ \\
     & & & & & & \\
 RMSE     & $40.81_{+21.45}^{+62.51}$  & $37.17_{+19.22}^{+60.95}$   &  $36.25_{+17.62}^{+50.78}$  & $34.27_{+16.54}^{+55.82}$  & $30.06_{+13.45}^{+53.29}$ & $31.80_{+14.20}^{+49.67}$  \\
     & & & & & & \\
 \hline
 \end{tabular}
\end{table*}
%
The very good model performances for the absolute temperature and for the wind direction
deduced from the overall statistical analysis are confirmed here (Tables~\ref{tab:br_temp_cumdist},
\ref{tab:br_ws_cumdist}, \ref{tab:br_ws_cumdist_5dom} and \ref{tab:br_wd_cumdist}).
The largest median bias for the temperature is found at Cerro Armazones, at 2~m, and is only 0.64$^o$C. The median RMSE is always good, with values
always inferior to 1$^o$C. Even more remarkable is the fact that 75$\%$ of the times, the RMSE is inferior to 1.23$^o$C at 2~m at
Cerro Armazones (worst value) and inferior to 0.90$^o$C at 30~m at Cerro Paranal (best value).\\ \\
\noindent
The wind direction median bias is very close to zero (between -5.45$^{\circ}$ and 7.28$^{\circ}$ at all levels), and the median RMSE
is in the interval [37.57$^{\circ}$ - 46.16$^{\circ}$], that corresponds to a $RMSE_{relative}$ in the range [21\% - 26\%].
What does it occur if we filter out the winds inferior to a given threshold (here 3~m$\cdot$s$^{-1}$), like in Section~\ref{sec:wd_ws}?
The results of this filtering are reported in Table~\ref{tab:br_wd_cumdist_filter}.
The median bias remain similar (between 1.01$^{\circ}$ and 8.55$^{\circ}$, in absolute values). However, the median RMSE 
are strongly reduced, and are now in the range [30.06-40.81$^{\circ}$], which correspond to a RMSE$_{relative}$ in the range [17-23\%].\\ \\
\noindent
Concerning the wind speed, the same conclusions obtained for the overall statistical analysis can be drawn, but the values of the
median bias and RMSE are slightly better.
When the highest horizontal resolution is $\Delta$X~=~500 m, the median bias ranges from 0.82~m$\cdot$s$^{-1}$ at 30~m at Cerro Paranal to 
2.40~m$\cdot$s$^{-1}$ at 2~m at Cerro Armazones.
The largest median RMSE is found at Cerro Armazones at 2~m and is equal to 2.71~m$\cdot$s$^{-1}$.
Performances are significantly improved when the highest model horizontal resolution is $\Delta$X~=~100~m. The median bias is halved, or better:
its best value is found at Cerro Armazones at 28~m and is equal to 0.12~m$\cdot$s$^{-1}$, and its worst value remains inferior to 1~m$\cdot$s$^{-1}$
(0.93~m$\cdot$s$^{-1}$ at both 2~m and 11~m high at Cerro Armazones).
Also, the improvement in the RMSE is not negligible.
Even though the RMSE median values are ranging
from 1.85~m$\cdot$s$^{-1}$ at Cerro Paranal at 2~m to 2.18~m$\cdot$s$^{-1}$ at 20~m at Cerro Armazones, i.e very
similar to the $\Delta$X~=~500~m configuration, the improvement is visible looking at the 1st and 3rd quartiles.
The values of the 3rd quartiles are strongly reduced. 
Its largest value with the $\Delta$X~=~500~m configuration was 6.41~m$\cdot$s$^{-1}$
(at 11~m at Cerro Armazones), it is now 3.34~m$\cdot$s$^{-1}$ (at 2~m at Cerro Armazones) with the $\Delta$X~=~100~m configuration.\\ \\
As we have seen, bias and RMSE of the night by night analysis are slightly better than the bias and RMSE of the overall statistics.
\section{Correlation coefficient}
\label{sec:cc}
In this section we explain why the use of the correlation coefficient isn't useful for our specific study, and 
why we do not suggest its use for such a statistical analysis.
The correlation coefficient is defined as:
\begin{equation}
cc=\frac{\sum_{i=1}^{N}(X_i-\overline{X})(Y_i-\overline{Y})}{\sqrt{\sum_{i=1}^{N}(X_i-\overline{X})^2}\sqrt{\sum_{i=1}^{N}(Y_i-\overline{Y})^2}}
\label{eq:cc}
\end{equation}
where $X_i$ are the individual observations, $Y_i$ the individual simulations parameters calculated at the same time and $N$ is
the number of times for which a couple ($X_i$,$Y_i$) is available with both $X_i$ and $Y_i$ different from zero.
\par
\begin{figure*}
\begin{center}
\begin{tabular}{c}
\includegraphics[width=0.75\textwidth]{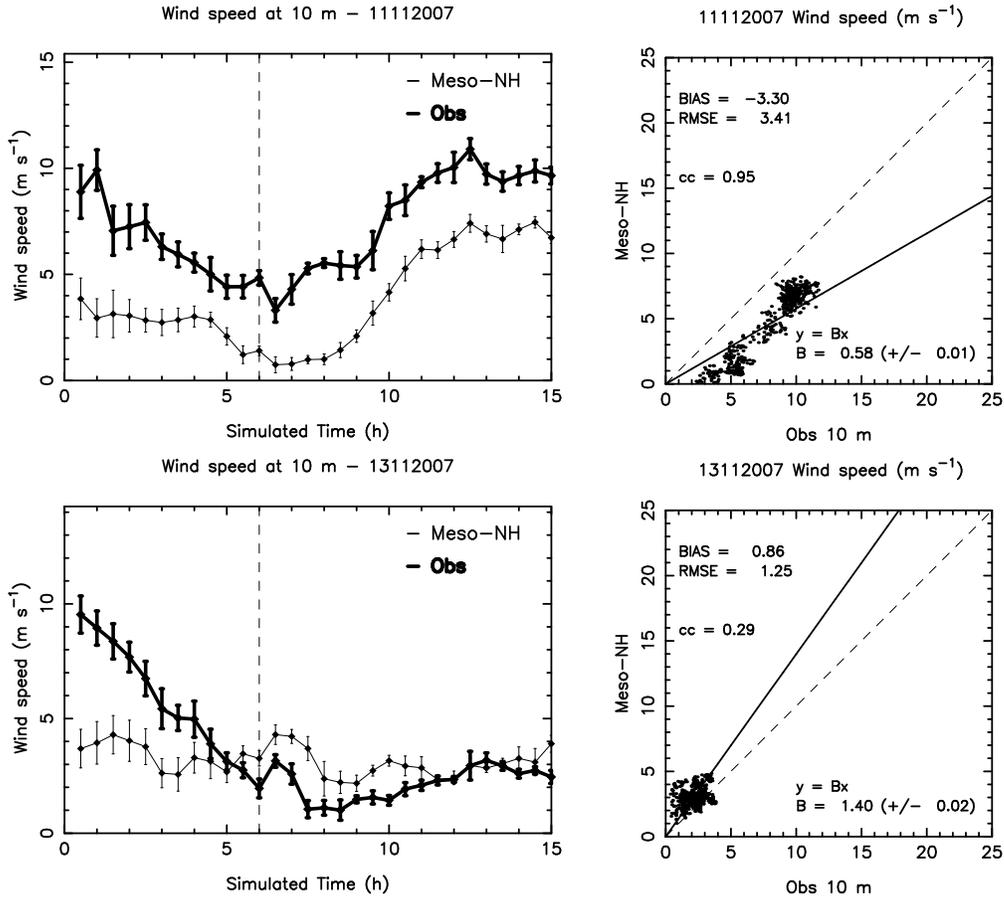}\\
\end{tabular}
\end{center}
\caption[cc_Ex]{\label{fig:cc_ex} Examples of the temporal evolution of the wind speed during 2 different nights (on the left) and the corresponding scattered plots with the bias, RMSE and $cc$ values (on the right). }
\end{figure*}

An analysis on the correlation coefficient put into evidence that it 
{\bf (1)} doesn't provide any further information with respect to the bias and the RMSE on the reliability of the model in reconstructing the
atmospherical parameters.
It doesn't provide any information on the forecast of the trend (as could do a simple temporal evolution average) since it doesn't take into account the time;
{\bf (2)} can be misleading in many cases (bad value for visibly good forecasts, and {\em vice-versa}).
This is particularly evident when we consider the night by night case.
An example is visible in Fig.~\ref{fig:cc_ex}.
We observe that the wind speed matches in a much better reliable way to observations during the night of 13~November than the night of 11~November
 while the $cc$ for the 11~November night is much better than for the 13~November night (0.95 vs. 0.29). 
%
\section{Conclusions}
This paper is the second of a series of articles aiming at demonstrating the Meso-Nh model ability in reconstructing
the  meteorological parameters and the optical turbulence.
Using a set of 20 nights in 2007 for which meteorological observations near the surface were available, we have deeply analysed the performances of the Meso-Nh
model in predicting the main meteorological parameters in the surface layer (wind speed, wind direction and temperature, from 2~m a. g. l. up to 30~m a. g. l. 
at two ESO sites (Cerro Paranal and Cerro Armazones).
Two different configurations were used, one made with 3 domains and the innermost domain horizontal mesh-size $\Delta$X~=~500~m, and the second one 
made with 5 domains and the innermost domain horizontal mesh-size $\Delta$X~=~100~m.
The higher model horizontal resolution $\Delta$X~=~100~m was employed in order to investigate its impact on the wind speed forecast near the ground 
(a higher horizontal resolution allows for a better description of the orography).
Statistical parameters as bias and RMSE have been used to quantify the model performances.
Here are the main results concerning the individual nights performances: \\
{\bf - Temperature}: minimum median bias is 0.03$^{\circ}$C (30~m, Cerro Paranal) and maximum median bias is 0.64$^{\circ}$C (2~m, Cerro Armazones);
minimum median RMSE is 0.64$^{\circ}$C (30~m, Cerro Paranal) and maximum median RMSE is 0.93$^{\circ}$C (30~m, Cerro Armazones);  \\
{\bf - Wind direction}: minimum median bias is 1.01$^{\circ}$ (10~m, Cerro Paranal) and maximum median bias is 7.28$^{\circ}$ (2~m, Cerro Armazones); 
minimum median RMSE is 37.57$^{\circ}$ (20~m, Cerro Armazones) and maximum median RMSE is 46.16$^{\circ}$ (30~m, Cerro Paranal); the RMSE$_{relative}$ 
is in the range [2\% - 26\%];\\
{\bf - Wind speed} ($\Delta$X~=~100~m configuration): minimum median bias is 0.12~m$\cdot$s$^{-1}$ (30~m, Cerro Armazones) and maximum median bias is 
0.93~m$\cdot$s$^{-1}$ (11~m, Cerro Armazones); minimum median RMSE is 1.85~m$\cdot$s$^{-1}$ (10~m, Cerro Paranal) and maximum median RMSE is 
2.18~m$\cdot$s$^{-1}$ (20~m, Cerro Armazones); \\ \\
\noindent
Concerning the wind direction, it is worth noting that the results can be strongly improved if we filter out the weak winds (inferior to 3~m$\cdot$s$^{-1}$).
On the individual nights, the median RMSE is reduced from the range [37.57-46.16$^{\circ}$] 
to the range [30.06-40.81$^{\circ}$] (median RMSE$_{relative}$ from [21-26\%] to [17-23\%]).
For the whole sample, the RMSE improves of around 15$^{\circ}$ (the
largest RMSE goes from 58$^{\circ}$ to 41.5$^{\circ}$, at 10~m at Cerro Paranal, for example).
The RMSE (with the filter applied) is reduced from the range [42.9$^{\circ}$ - 58$^{\circ}$] to the range [31.7$^{\circ}$ - 43$^{\circ}$].  
The corresponding RMSE$_{relative}$ is reduced from [24 - 32\%] to [18 - 24\%].
We can say that the strongest is the wind, the more accurate is the model reconstruction of the wind direction. 
This is a very positive result, because for an astronomer using ground-based facilities it is more 
important to know the wind direction accurately when the wind is strong.
As we already anticipated, the wind direction is a parameter that is mostly 
correlated to the seeing conditions above Observatories. 
Moreover, experience showed that the useful information in astronomical context is 
typically the identification of the quadrant from which the wind comes from.
A median RMSE$_{relative}$ after filtering of [17-23\%] (associated to a median RMSE of [30.06 - 40.81$^{\circ}$])
appears therefore as a satisfactory result for the kind of application we are dealing with.  \\
It is planned by the MOSE team to investigate also the wind direction using the $\Delta$X~=~100~m configuration, with expected even better bias and RMSE, 
since the orography is better reconstructed and allows for a better description of the flow near the surface, when a higher horizontal resolution 
is employed. \\ 
It is also planned to increase the statistical sample in order to make the conclusions even more robust.\\ \\
\noindent
In conclusion, we have demonstrated that the best results in terms of model forecast efficiency 
are for the absolute temperature and the wind direction, with biases close to zero and small RMSEs.
The results for the wind speed are satisfactory too, even though it seems that a higher horizontal resolution is needed to achieve satisfactory performances. 
\noindent
At these levels of accuracy, the forecasted meteorological parameters can reveal very useful for astronomers using ground-based facilities. 
For example, one could think of using the forecasted temperature of the beginning of the night for the thermalization of the dome of the telescope.
More over, knowing in advance the main direction of the wind near the surface could allow the astronomer to anticipate the occurrence of a good/bad seeing night, 
and plan the observations consequently. 
All the different applications that can be derived 
from such good forecasts, and the context in which they apply, are discussed more in detail in the joint paper (Masciadri et al., 2013). \\ \\
\noindent
In a forthcoming paper we will show that, using further quantitative estimators
of the model performances and employing dedicated at posteriori factors of correction, it is possible to 
prove the usefulness of the model for the wind speed also using a horizontal resolution $\Delta$X~=~500~m.
\section*{Acknowledgements}
This study is co-funded by the ESO contract: E-SOW-ESO-245-0933 (MOSE Project).
Meteorological data-set from the Automatic Weather Station (AWS) and mast at Cerro Armazones are from the Thirty Meter Telescope Site Testing -
Public Database Server \citep{schoeck09}.
Meteorological data-set from the AWS and mast at Cerro Paranal are from ESO Astronomical Site Monitor (ASM - Doc.N. VLT-MAN-ESO-17440-1773). 
We are very grateful to the whole staff of the TMT Site Testing Working Group for providing information about their data-set as well as 
to the ESO Board of MOSE (Marc Sarazin, Pierre-Yves Madec, Florian Kerber and Harald Kuntschner) for their constant support to this study. We acknowledge 
M. Sarazin and F. Kerber for providing us the ESO data-set used in this study.  
A great part of simulations are run on the HPCF cluster of the European Centre for Medium Weather Forecasts (ECMWF) - Project SPITFOT.
%

\appendix
\noindent
\section{Temporal evolutions}
We report here the temporal evolution of the average, bias and RMSE for all the atmospherical parameters (wind speed and direction, and temperature),
at Cerro Paranal and Cerro Armazones.
\begin{figure*}
\begin{center}
\begin{tabular}{c}
\includegraphics[width=\textwidth]{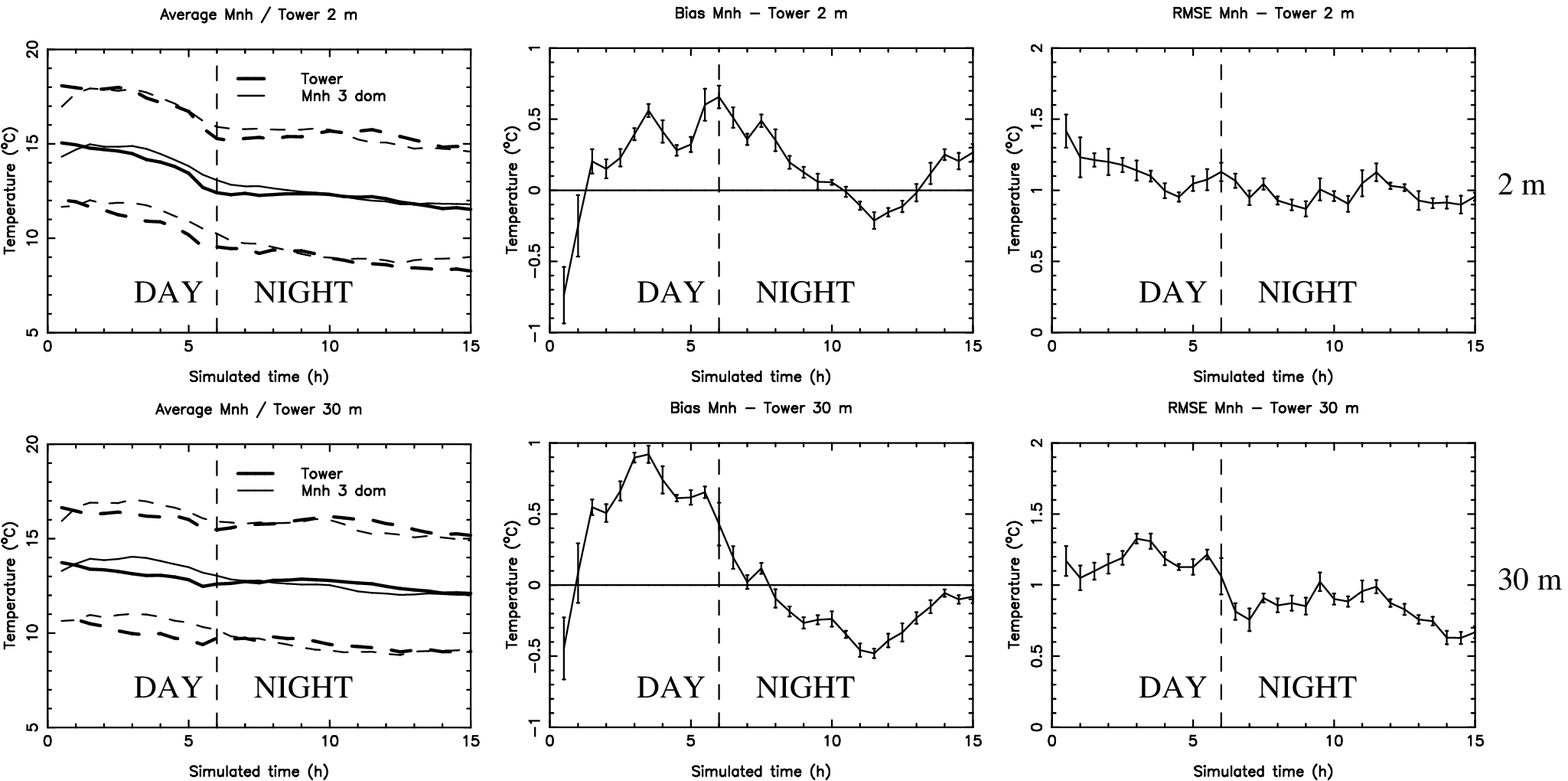}
\end{tabular}
\end{center}
\caption[Surface_temp_paranal_evol]{\label{fig:Surface_temp_paranal_evol} Temporal evolution of the absolute temperature average (the bold line is the
observation average and the thin line is Meso-Nh average), bias (Mnh - Observations)
and RMSE at Cerro Paranal (top: at 2~m; bottom: at 30~m).
The x-axis represents the time from the beginning of the simulation (00 h is 18 UT of the day before, 06 h is 00 UT,
and 15 h is 09 UT). The nights starts at around 06 h (00 UT / 20 LT) and is delimited by the vertical dashed line.
As explained in the text the model configuration selected for this study is optimized for the investigation of the night period.
Meso-Nh is with the $\Delta$X~=~500 m configuration.
Error bars are $\pm \sigma$ (standard deviation) for each of the 30-minutes intervals.}
\end{figure*}
\begin{figure*}
\begin{center}
\begin{tabular}{c}
\includegraphics[width=\textwidth]{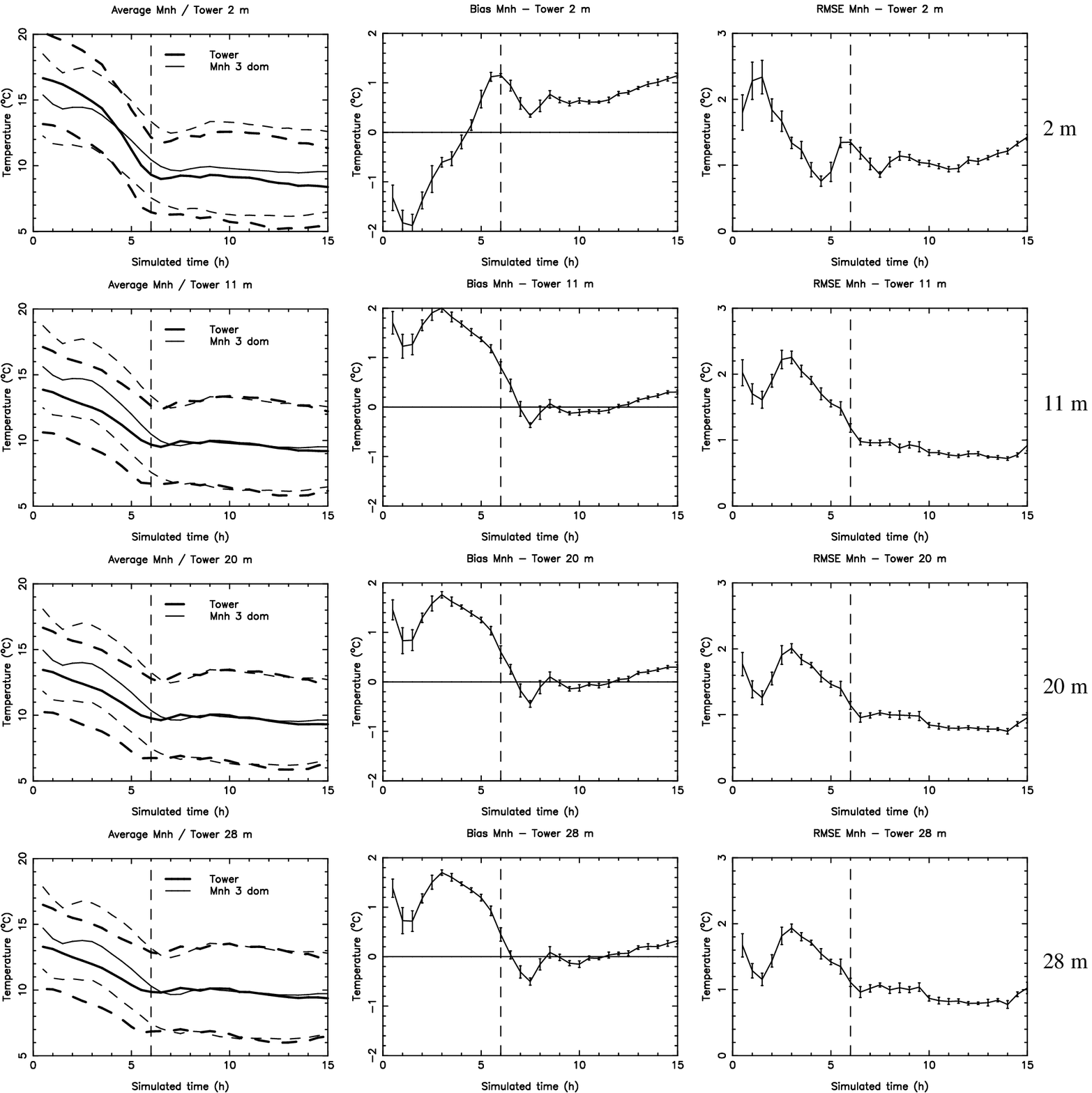}
\end{tabular}
\end{center}
\caption[Surface_temp_armazones_evol]{\label{fig:Surface_temp_armazones_evol} Same as Fig.~\ref{fig:Surface_temp_paranal_evol} but for the 
absolute temperature at Cerro Armazones (from top to bottom: at 2~m, 11~m, 20~m and 28~m).}
\end{figure*}
\begin{figure*}
\begin{center}
\begin{tabular}{c}
\includegraphics[width=\textwidth]{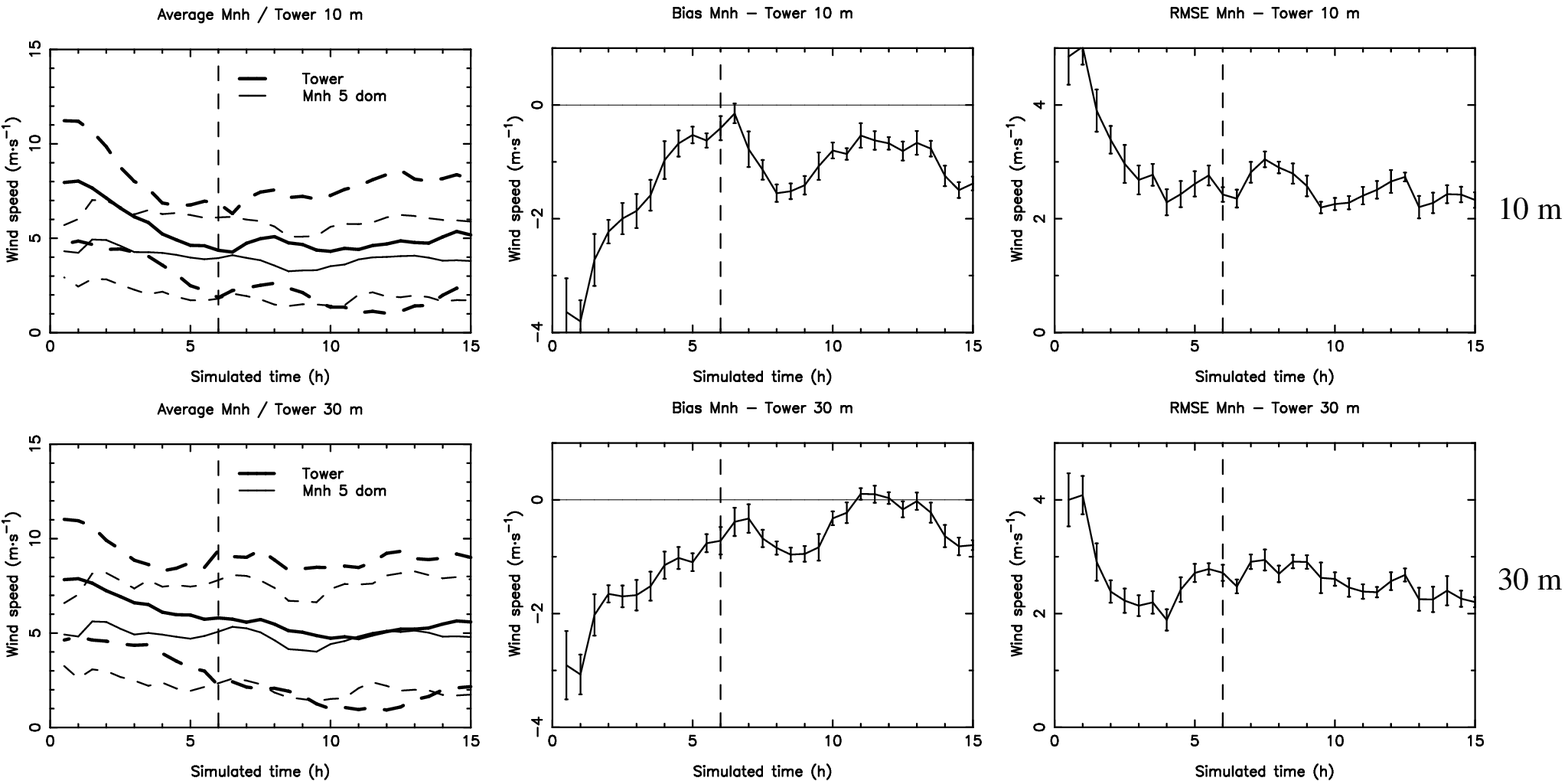}
\end{tabular}
\end{center}
\caption[Surface_ws_paranal_evol]{\label{fig:Surface_ws_paranal_evol} Same as Fig.~\ref{fig:Surface_temp_paranal_evol} but for
 the wind speed
 at Cerro Paranal (top: at 10~m; bottom: at 30~m). Meso-Nh
is used with $\Delta$X~=~100~m configuration.}
\end{figure*}
\begin{figure*}
\begin{center}
\begin{tabular}{c}
\includegraphics[width=\textwidth]{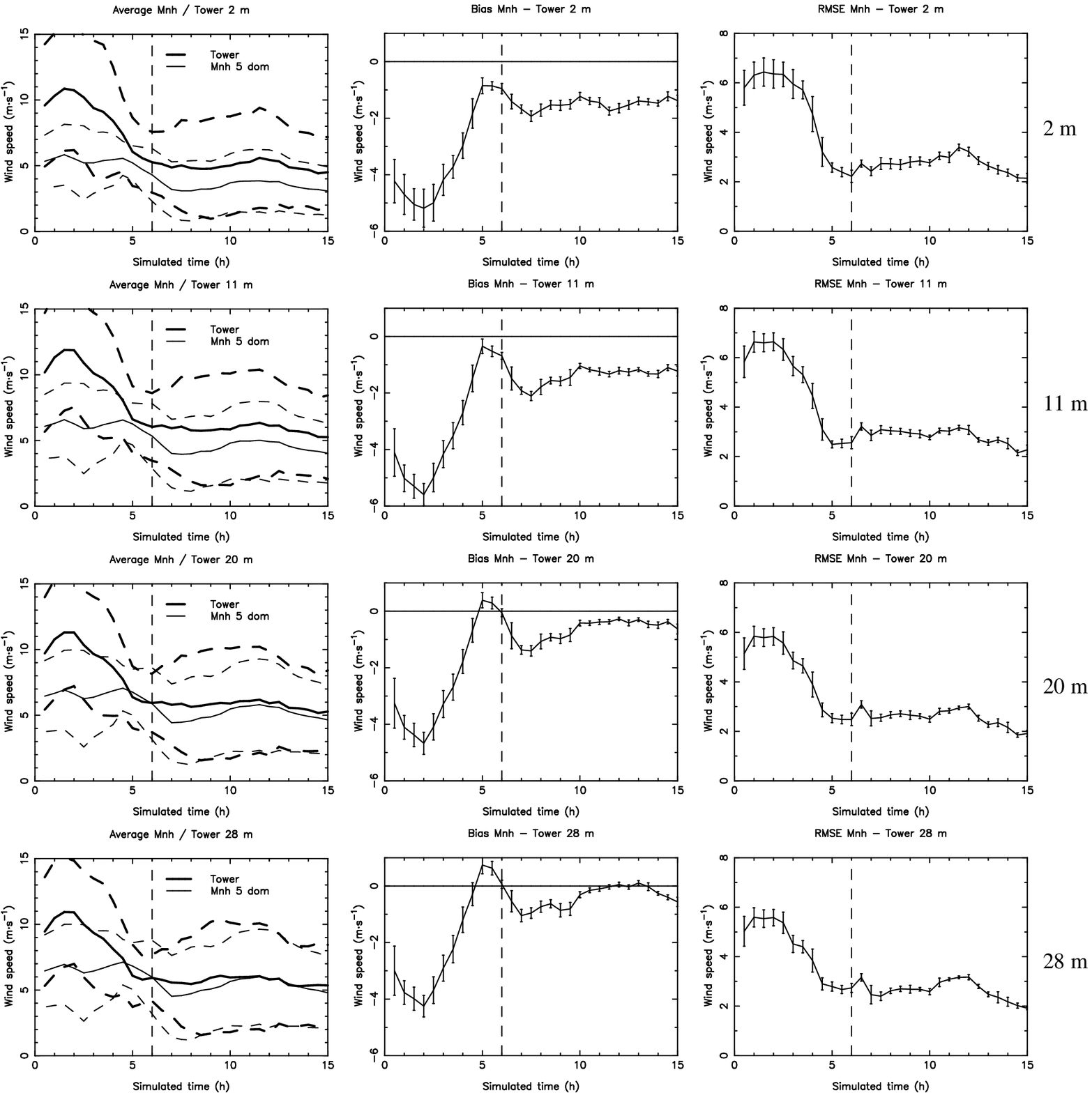}
\end{tabular}
\end{center}
\caption[Surface_ws_armazones_evol]{\label{fig:Surface_ws_armazones_evol} Same as Fig.~\ref{fig:Surface_temp_paranal_evol} but for
 the wind speed
 at Cerro Armazones (from top to bottom: at 2~m, 11~m, 20~m and 28~m). Meso-Nh
is used with $\Delta$X~=~100~m configuration.}
\end{figure*}
\begin{figure*}
\begin{center}
\begin{tabular}{c}
\includegraphics[width=\textwidth]{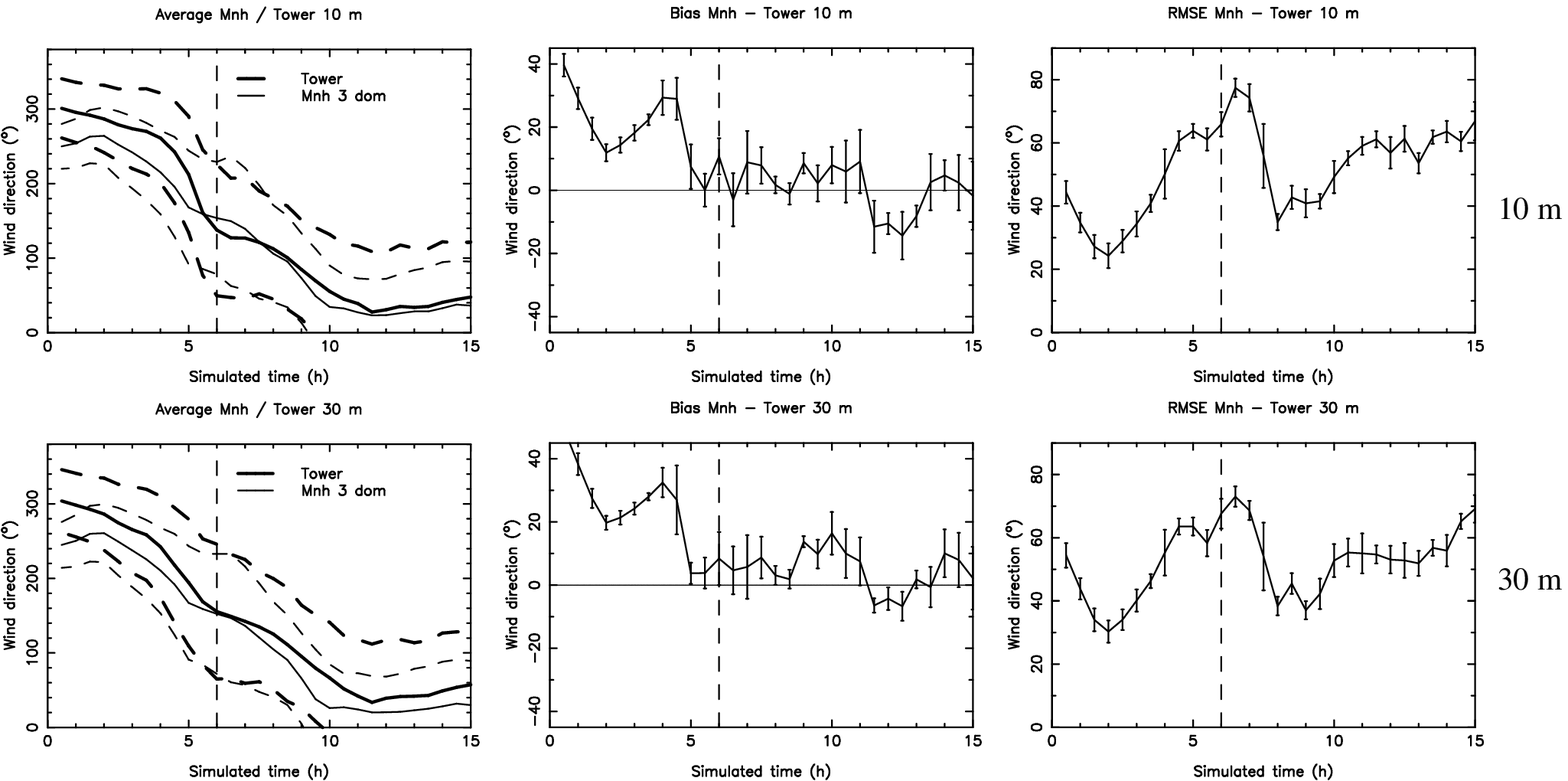}
\end{tabular}
\end{center}
\caption[Surface_wd_paranal_evol]{\label{fig:Surface_wd_paranal_evol} Same as Fig.~\ref{fig:Surface_temp_paranal_evol} but for
 the wind direction
 at Cerro Paranal (top: at 10~m; bottom: at 30~m).}
\end{figure*}
\begin{figure*}
\begin{center}
\begin{tabular}{c}
\includegraphics[width=\textwidth]{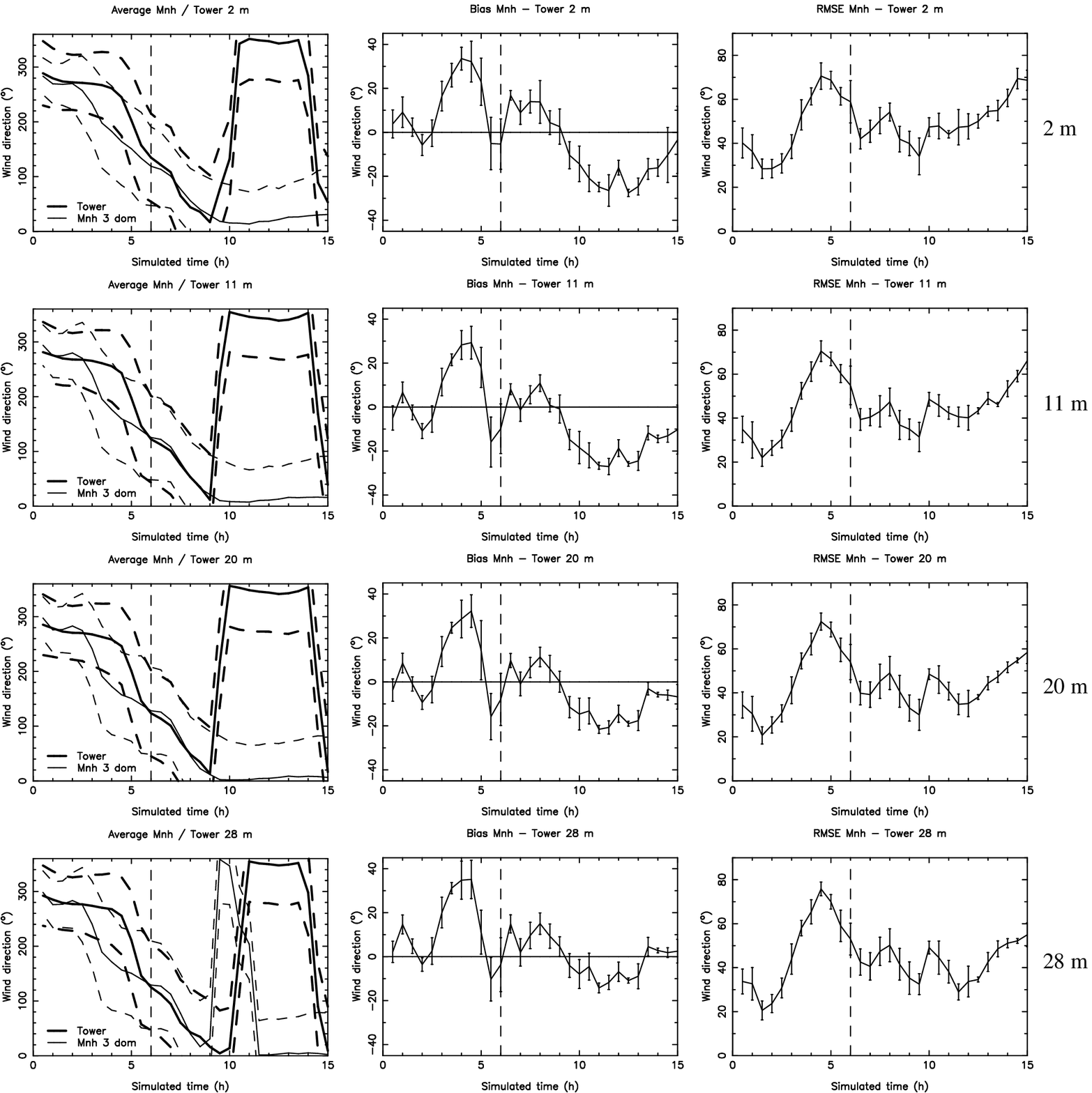}
\end{tabular}
\end{center}
\caption[Surface_wd_armazones_evol]{\label{fig:Surface_wd_armazones_evol} Same as Fig.~\ref{fig:Surface_temp_paranal_evol} but for
 the wind direction
 at Cerro Armazones (from top to bottom: at 2~m, 11~m, 20~m and 28~m).}
\end{figure*}

\section{Individual nights model performances: cumulative distributions}
We report here all the cumulative distributions of the bias and RMSE of the single nights or all the atmospherical parameters (wind speed and direction, 
and temperature), at Cerro Paranal and Cerro Armazones
\begin{figure*}
\begin{center}
\begin{tabular}{c}
\includegraphics[width=0.70\textwidth]{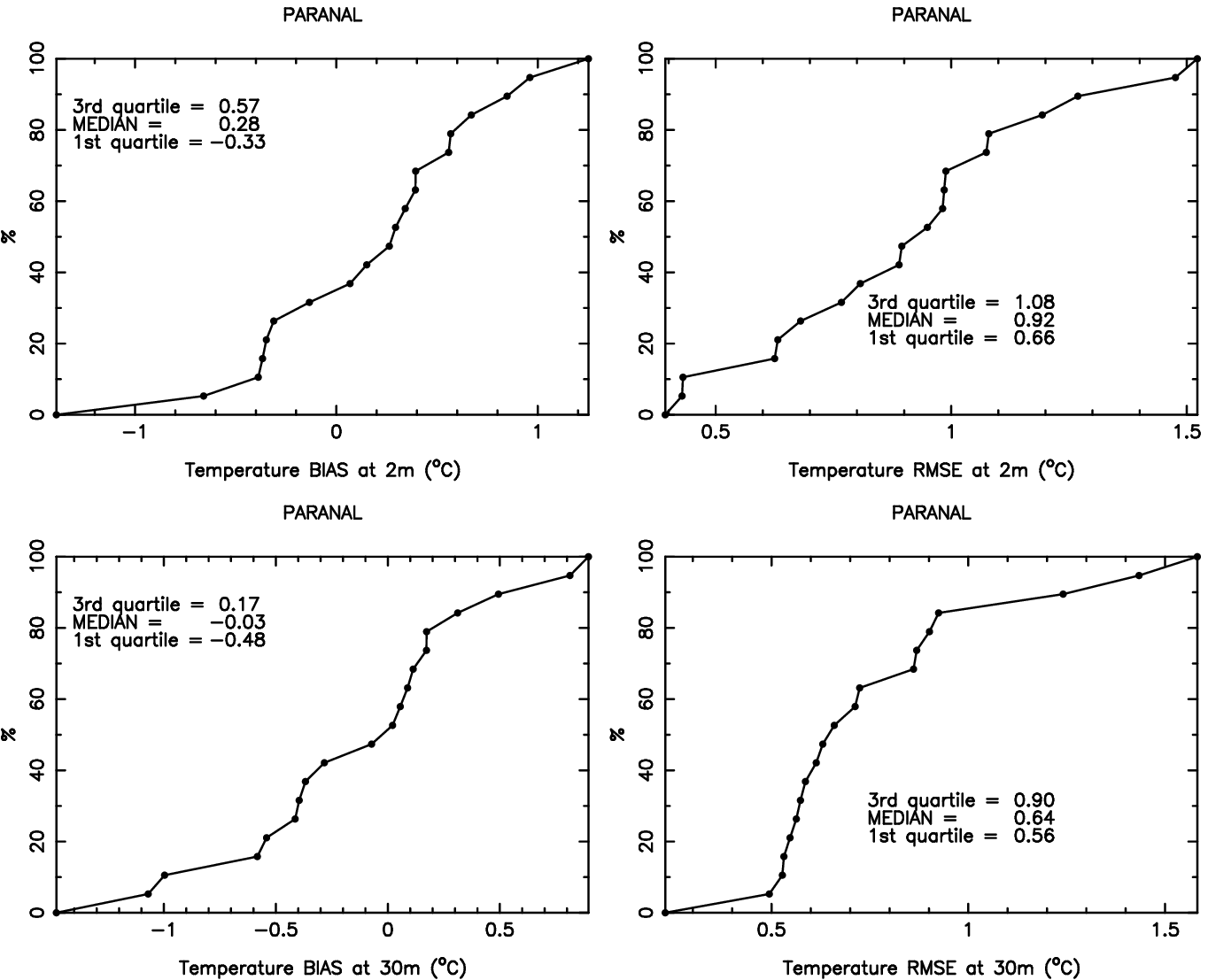}
\end{tabular}
\end{center}
\caption[Surface_temp_paranal_cumdist]{\label{fig:Surface_temp_paranal_cumdist} Cumulative distribution of bias (Mnh - Observations, on the left)
and RMSE 
(on the right)
of the temperature at Cerro Paranal (20 nights sample), at 2~m (top) and 30~m (bottom).
See Table~\ref{tab:br_temp_cumdist} for a
summarize of the bias and RMSE values. Meso-Nh is in the $\Delta$X~=~500~m configuration.}
\end{figure*}
\begin{figure*}
\begin{center}
\begin{tabular}{c}
\includegraphics[width=0.70\textwidth]{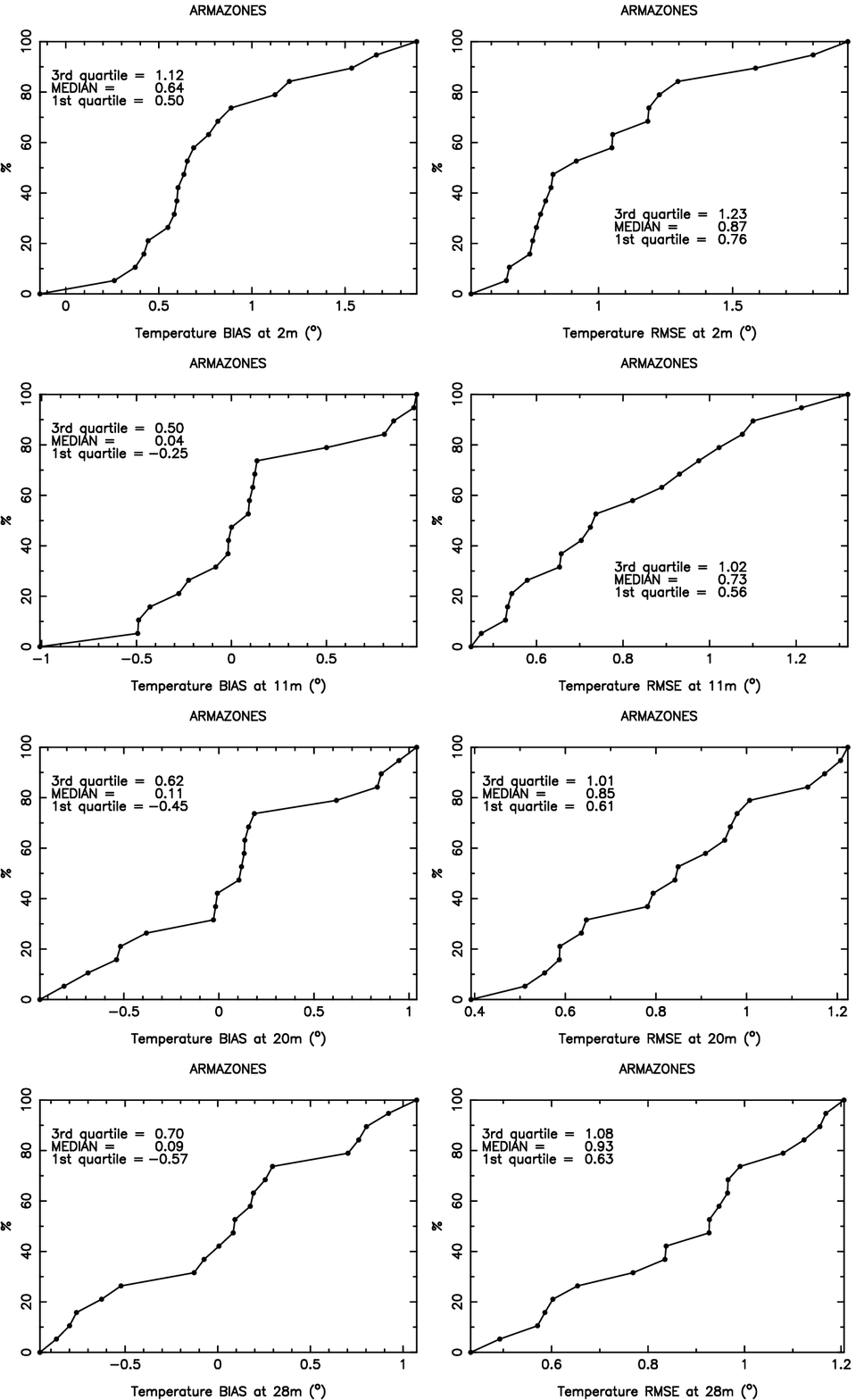}
\end{tabular}
\end{center}
\caption[Surface_temp_armazones_cumdist]{\label{fig:Surface_temp_armazones_cumdist} Cumulative distribution of bias (Mnh - Observations, on the left)
and RMSE 
(on the right)
of the temperature at Cerro Armazones (20 nights sample), at 2~m, 11~m, 20~m and 28~m (from top to bottom, respectively).
See Table~\ref{tab:br_temp_cumdist} for a
summarize of the bias and RMSE values. Meso-Nh is in the $\Delta$X~=~500~m configuration.}
\end{figure*}
\begin{figure*}
\begin{center}
\begin{tabular}{c}
\includegraphics[width=0.70\textwidth]{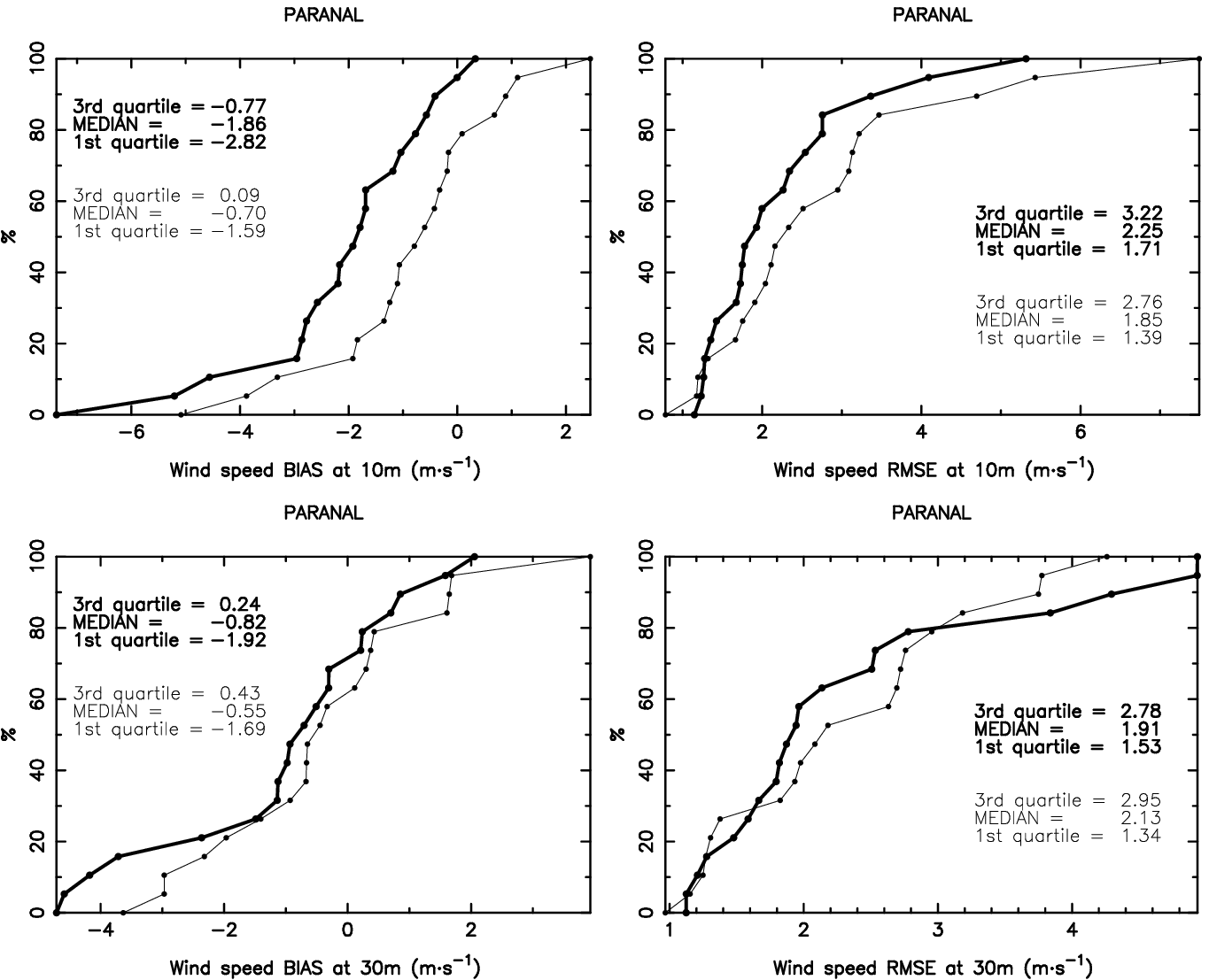}
\end{tabular}
\end{center}
\caption[Surface_ws_paranal_cumdist]{\label{fig:Surface_ws_paranal_cumdist} Cumulative distribution of bias (Mnh - Observations, on the left)
and RMSE 
(on the right)
of the wind speed at Cerro Paranal (20 nights sample), at 10~m (top) and 30~m (bottom).
In thick line, Meso-Nh is with the standard horizontal
resolution ($\Delta$X~=~500 m). In thin line, with the high horizontal resolution ($\Delta$X~=~100 m). See Tables \ref{tab:br_ws_cumdist}
and \ref{tab:br_ws_cumdist_5dom} for a summarize of the bias and RMSE values.}
\end{figure*}
\begin{figure*}
\begin{center}
\begin{tabular}{c}
\includegraphics[width=0.70\textwidth]{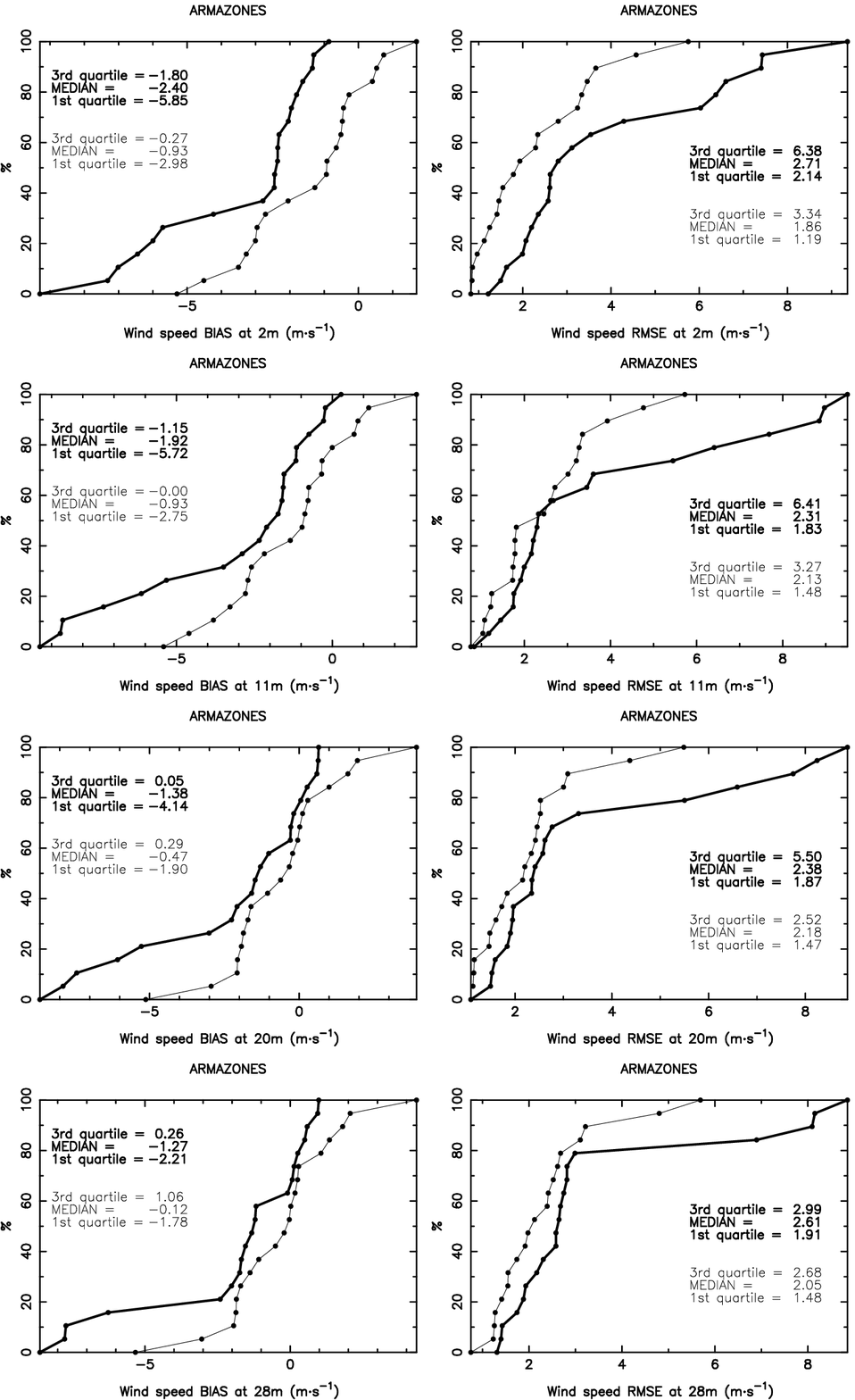}
\end{tabular}
\end{center}
\caption[Surface_ws_armazones_cumdist]{\label{fig:Surface_ws_armazones_cumdist} Cumulative distribution of bias (Mnh - Observations, on the left)
and RMSE 
(on the right)
of the wind speed at Cerro Armazones (20 nights sample), at 2~m, 11~m, 20~m and 28~m (from top to bottom, respectively).
See Tables \ref{tab:br_ws_cumdist}
and \ref{tab:br_ws_cumdist_5dom} for a summarize of the bias and RMSE values.
In thick line, Meso-Nh is with the standard horizontal
resolution ($\Delta$X~=~500 m). In thin line, with the high horizontal resolution ($\Delta$X~=~100 m).}
\end{figure*}
\begin{figure*}
\begin{center}
\begin{tabular}{c}
\includegraphics[width=0.70\textwidth]{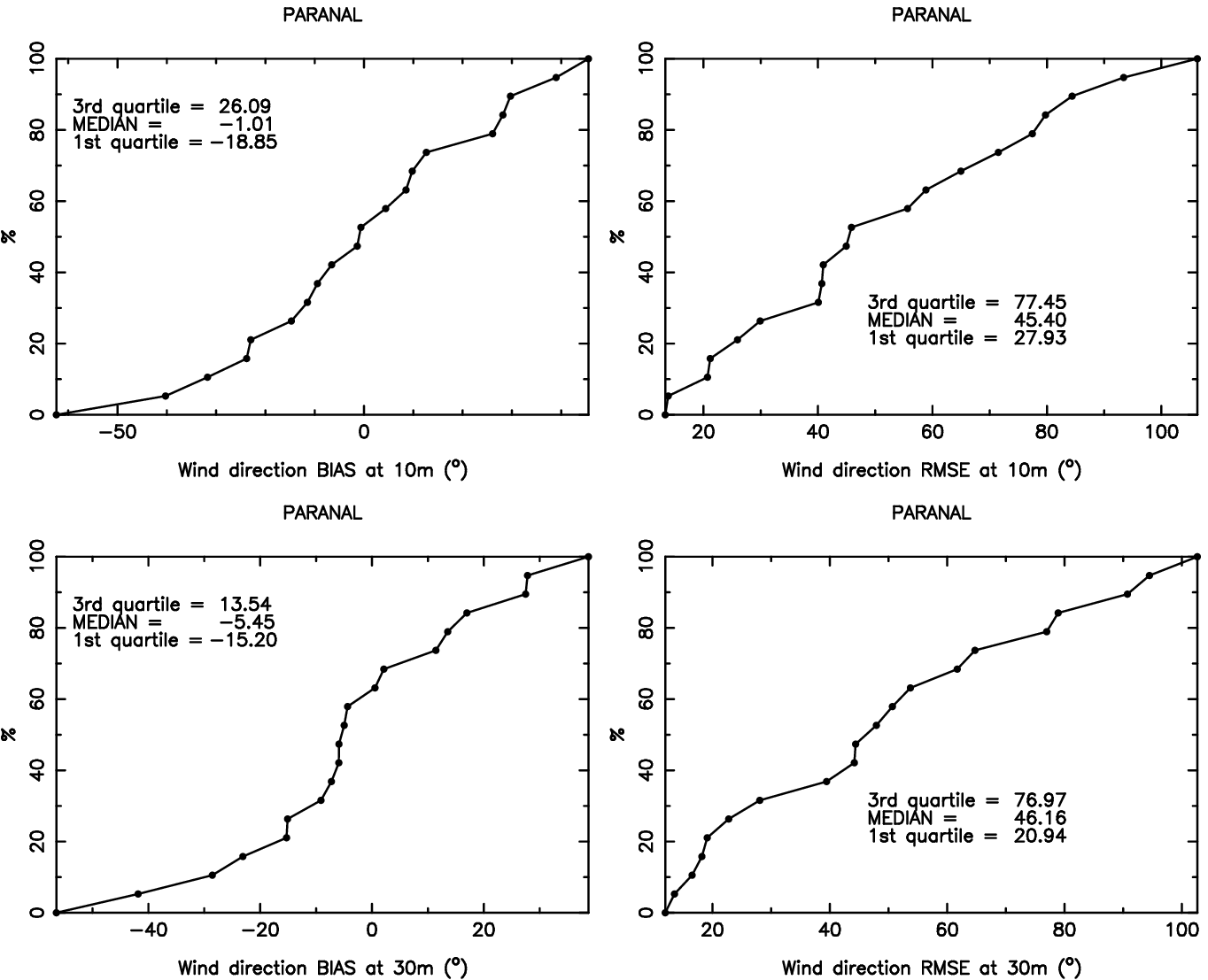}
\end{tabular}
\end{center}
\caption[Surface_wd_paranal_cumdist]{\label{fig:Surface_wd_paranal_cumdist} Cumulative distribution of bias (Mnh - Observations, on the left)
and RMSE 
(on the right)
of the wind direction at Cerro Paranal (20 nights sample), at 10~m (top) and 30~m (bottom).
See Table~\ref{tab:br_wd_cumdist}
 for a summarize of the bias and RMSE values. Meso-Nh is with the standard horizontal
resolution ($\Delta$X~=~500 m).}
\end{figure*}
\begin{figure*}
\begin{center}
\begin{tabular}{c}
\includegraphics[width=0.70\textwidth]{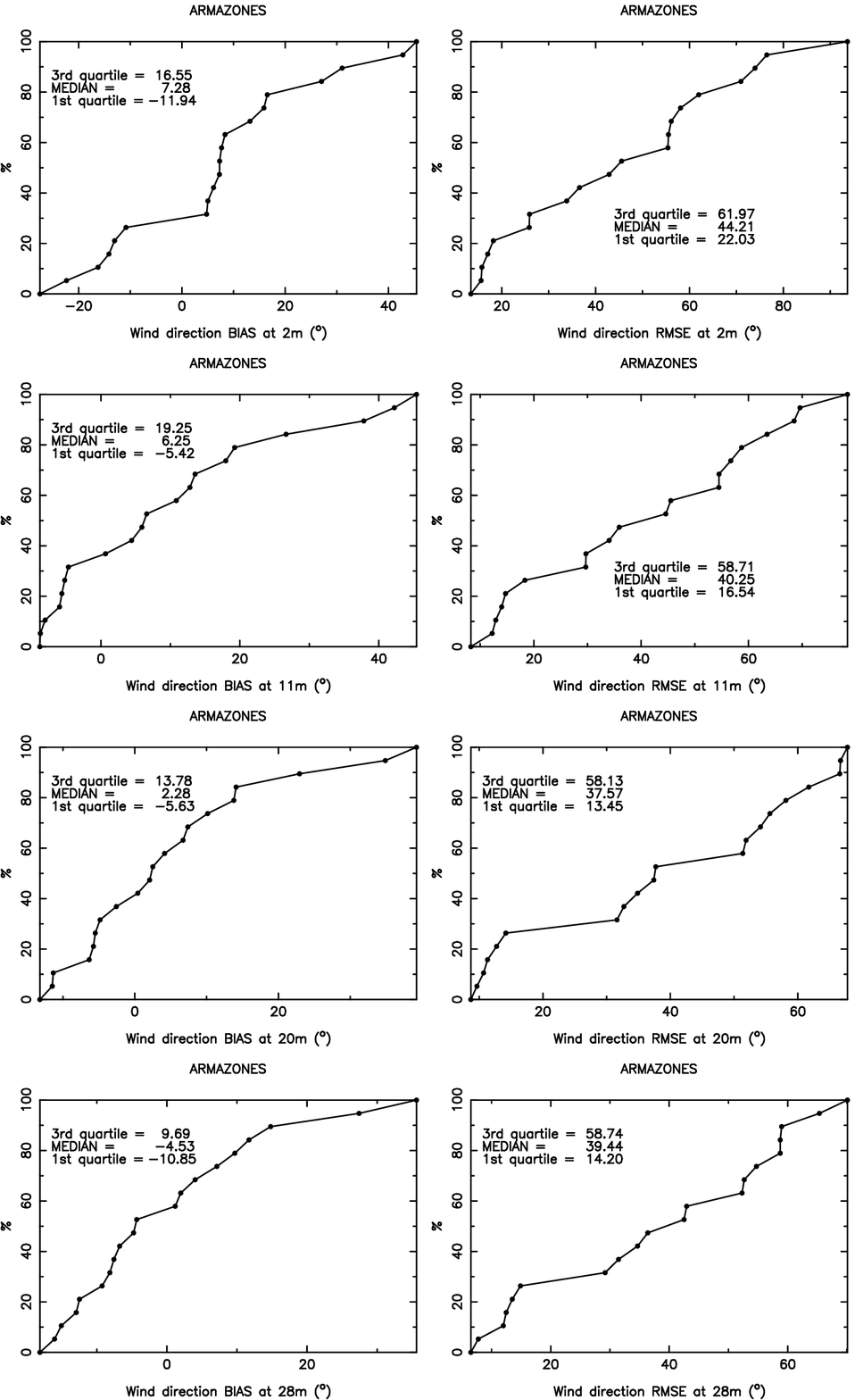}
\end{tabular}
\end{center}
\caption[Surface_wd_armazones_cumdist]{\label{fig:Surface_wd_armazones_cumdist} Cumulative distribution of bias (Mnh - Observations, on the left)
and RMSE 
(on the right)
of the wind direction at Cerro Armazones (20 nights sample), at 2~m, 11~m, 20~m and 28~m (from top to bottom, respectively). See Table~\ref{tab:br_wd_cumdist}
 for a summarize of the bias and RMSE values.
Meso-Nh is with the standard horizontal
resolution ($\Delta$X~=~500 m). }
\end{figure*}

\label{lastpage}

\begin{thebibliography}{}
%
\bibitem[\protect\citeauthoryear{Dali Ali et al.}{2010}]{dali10}
Dali Ali W. et al., 2010, A\&A, 524, id.A73
\bibitem[\protect\citeauthoryear{Davies \& Thomson}{1999}]{davies99}
Davies B. M., Thomson D. J., 1999, Atmos. Environ., 33, 4909
\bibitem[\protect\citeauthoryear{Fisher \&  Lee}{1983}]{fisher83}
Fisher N. I., Lee A. J., 1983, Biometrika, 70, 327
\bibitem[\protect\citeauthoryear{Hagelin, Masciadri \&  Lascaux}{2010}]{hagelin10}
Hagelin S., Masciadri E., Lascaux F., 2010, MNRAS, 407, 2230
\bibitem[\protect\citeauthoryear{Hagelin, Masciadri \&  Lascaux}{2011}]{hagelin11}
Hagelin S., Masciadri E., Lascaux F., 2011, MNRAS, 412, 2695
\bibitem[\protect\citeauthoryear{Jim\'enez \& Dudhia}{2013}]{jimenez13}
Jim\'enez P. A., Dudhia, J., 2013, J. Appl. Meteor. Climatol., 52, 1610
\bibitem[\protect\citeauthoryear{Joffre \& Laurila}{1988}]{joffre88}
Joffre S., Laurila T., 1988, J. Appl. Meteor, 27, 550
\bibitem[\protect\citeauthoryear{Lafore et al.}{1998}]{lafore98}
Lafore J.-P. et al.,
1998, Annales Geophysicae, 16, 90
\bibitem[\protect\citeauthoryear{Lascaux et al.}{2009}]{lascaux09}
Lascaux F., Masciadri E., Hagelin S., Stoesz J., 2009, MNRAS, 398, 1093
\bibitem[\protect\citeauthoryear{Mahrt}{2011}]{mahrt11}
Mahrt L., 2011, J. Appl. Meteor. Climatol., 50, 144
\bibitem[\protect\citeauthoryear{Masciadri \& Garfias}{2001}]{masciadri01a}
Masciadri E., Garfias T., 2001, A\&A,  366, 708
\bibitem[\protect\citeauthoryear{Masciadri, Vernin \& Bougeault}{2001}]{masciadri01b}
Masciadri E., Vernin J., Bougeault P., 2001, A\&A,  365, 699
\bibitem[\protect\citeauthoryear{Masciadri}{2003}]{masciadri03}
Masciadri E., 2003, RMxAA,  39, 249
\bibitem[\protect\citeauthoryear{Masciadri, Lascaux \& Fini}{2013}]{masciadri13}
Masciadri E., Lascaux F., Fini L., 2013, MNRAS, accepted for publication
\bibitem[\protect\citeauthoryear{Sandrock \& Amestica}{2009}]{sandrock99}
Sandrock S., Amestica R., 1999, Doc no.: VLT-MAN-ESO-17440-1773
\bibitem[\protect\citeauthoryear{Schoeck et al.}{2009}]{schoeck09}
Schoeck M. et al., 2009, PASP, 121, 384
\bibitem[\protect\citeauthoryear{Skidmore, Travouillon \&  Riddle }{2007}]{skidmore07}
Skidmore W., Travouillon T., Riddle R., 2007, Internal TMT Report
\bibitem[\protect\citeauthoryear{Stein et al.}{2000}]{stein00}
Stein J., Richard E., Lafore J.-P., Pinty J.-P., Asencio N., Cosma S.,
2000, Meteorol. Atmos. Phys., 72, 203
\end{thebibliography}
\end{document}